\documentstyle[12pt]{article}

\advance \topmargin by -\headheight
\advance \topmargin by -\headsep
     
\evensidemargin \oddsidemargin
\begin{document}
\newcommand{\sect}[1]{\setcounter{equation}{0}\section{#1}}
\renewcommand{\theequation}{\thesection.\arabic{equation}}
\date{}
\topmargin -.6in
\def\nonu{\nonumber}
\def\rf#1{(\ref{eq:#1})}
\def\lab#1{\label{eq:#1}} 
\def\br{\begin{eqnarray}}
\def\er{\end{eqnarray}}
\def\be{\begin{equation}}
\def\ee{\end{equation}}
\def\eq{\!\!\!\! &=& \!\!\!\! }
\def\0{\nonumber}
\def\lb{\lbrack}
\def\rb{\rbrack}
\def\({\left(}
\def\){\right)}
\def\v{\vert}
\def\bv{\bigm\vert}
\def\lskip{\vskip\baselineskip\vskip-\parskip\noindent}
\relax
\newcommand{\nit}{\noindent}
\newcommand{\ct}[1]{\cite{#1}}
\newcommand{\bi}[1]{\bibitem{#1}}
\def\a{\alpha}
\def\b{\beta}
\def\ca{{\cal A}}
\def\cm{{\cal M}}
\def\cn{{\cal N}}
\def\cf{{\cal F}}
\def\d{\delta}
\def\D{\Delta}
\def\eps{\epsilon}
\def\g{\gamma}
\def\G{\Gamma}
\def\grad{\nabla}
\def\h{ {1\over 2}  }
\def\hc{\hat{c}}
\def\hd{\hat{d}}
\def\hg{\hat{g}}
\def\hp{ {+{1\over 2}}  }
\def\hm{ {-{1\over 2}}  }
\def\k{\kappa}
\def\l{\lambda}
\def\L{\Lambda}
\def\lg{\langle}
\def\m{\mu}
\def\n{\nu}
\def\o{\over}
\def\om{\omega}
\def\O{\Omega}
\def\p{\phi}
\def\P{\Phi}
\def\pa{\partial}
\def\pr{\prime}
\def\ra{\rightarrow}
\def\rh{\rho}
\def\rg{\rangle}
\def\s{\sigma}
\def\t{\tau}
\def\th{\theta}
\def\Th{\Theta}
\def\ti{\tilde}
\def\wti{\widetilde}
\def\inte{\int dx }
\def\xb{\bar{x}}
\def\yb{\bar{y}}

\def\tr{\mathop{\rm tr}}
\def\Tr{\mathop{\rm Tr}}
\def\partder#1#2{{\partial #1\over\partial #2}}
\def\ds{{\cal D}_s}
\def\wtwo{{\wti W}_2}
\def\lie{{\cal G}}
\def\alie{{\widehat \lie}}
\def\dlie{{\cal G}^{\ast}}
\def\elie{{\widetilde \lie}}
\def\edlie{{\elie}^{\ast}}
\def\hlie{{\cal H}}
\def\wlie{{\widetilde \lie}}

\font \msb=msbm10 scaled \magstep1
\newcommand{\IR}{\mbox{\msb R} }
\newcommand{\IC}{\mbox{\msb C} }
\newcommand{\IP}{\mbox{\msb P} }
\newcommand{\IZ}{\mbox{\msb Z} }

\newfam\euffam
\font\sixeuf=eufm6
\font\eighteuf=eufm8
\font\twelveeuf=eufm10 scaled\magstep1
	\textfont\euffam=\twelveeuf
	\scriptfont\euffam=\eighteuf
  	\scriptscriptfont\euffam=\sixeuf
\def\euf{\fam\euffam\twelveeuf}
\def\sg{{\euf s}}
\def\ug{{\euf u}}
\newcommand\sbr[2]{\left\lbrack\,{#1}\, ,\,{#2}\,\right\rbrack} 
\newcommand\twomat[4]{\left(\begin{array}{cc}  
{#1} & {#2} \\ {#3} & {#4} \end{array} \right)}
\newcommand\twocol[2]{\left(\begin{array}{cc}  
{#1} \\ {#2} \end{array} \right)}
\newcommand\threemat[9]{\left(\begin{array}{ccc}  
{#1} & {#2} & {#3}\\ {#4} & {#5} & {#6}\\
{#7} & {#8} & {#9} \end{array} \right)}

\def\cA{{\cal A}}
\def\cB{{\cal B}}
\def\cC{{\cal C}}
\def\cD{{\cal D}}
\def\cE{{\cal E}}
\def\cF{{\cal F}}
\def\cG{{\cal G}}
\def\cH{{\cal H}}
\def\cI{{\cal I}}
\def\cJ{{\cal J}}
\def\cK{{\cal K}}
\def\cL{{\cal L}}
\def\cM{{\cal M}}
\def\cN{{\cal N}}
\def\cO{{\cal O}}
\def\cP{{\cal P}}
\def\cQ{{\cal Q}}
\def\cR{{\cal R}}
\def\cS{{\cal S}}
\def\cT{{\cal T}}
\def\cU{{\cal U}}
\def\cV{{\cal V}}
\def\cW{{\cal W}}
\def\cY{{\cal Y}}
\def\cZ{{\cal Z}}

\def\bfs{{\bf s}}
\def\bfts{{\bf {\tilde s}}}
\def\kere{\mbox{\rm Ker (ad $E$)}}
\def\ime{\mbox{\rm Im (ad $E$)}}
\def\qs{Q_{\bfs}}
\def\cgh{{\widehat {\cal G}}}
\newtheorem{definition}{Definition}[section]
\newtheorem{proposition}{Proposition}[section]
\newtheorem{theorem}{Theorem}[section]
\newtheorem{lemma}{Lemma}[section]
\newtheorem{corollary}{Corollary}[section]
\def\proof{\par{\it Proof}. \ignorespaces} \def\endproof{{$\Box$}\par}
\newenvironment{Proof}{\proof}{\endproof} 
%
\newcommand\PRL[3]{{\sl Phys. Rev. Lett.} {\bf#1} (#2) #3}
\newcommand\NPB[3]{{\sl Nucl. Phys.} {\bf B#1} (#2) #3}
\newcommand\NPBFS[4]{{\sl Nucl. Phys.} {\bf B#2} [FS#1] (#3) #4}
\newcommand\CMP[3]{{\sl Commun. Math. Phys.} {\bf #1} (#2) #3}
\newcommand\PRD[3]{{\sl Phys. Rev.} {\bf D#1} (#2) #3}
\newcommand\PLA[3]{{\sl Phys. Lett.} {\bf #1A} (#2) #3}
\newcommand\PLB[3]{{\sl Phys. Lett.} {\bf #1B} (#2) #3}
\newcommand\JMP[3]{{\sl J. Math. Phys.} {\bf #1} (#2) #3}
\newcommand\PTP[3]{{\sl Prog. Theor. Phys.} {\bf #1} (#2) #3}
\newcommand\SPTP[3]{{\sl Suppl. Prog. Theor. Phys.} {\bf #1} (#2) #3}
\newcommand\AoP[3]{{\sl Ann. of Phys.} {\bf #1} (#2) #3}
\newcommand\PNAS[3]{{\sl Proc. Natl. Acad. Sci. USA} {\bf #1} (#2) #3}
\newcommand\RMP[3]{{\sl Rev. Mod. Phys.} {\bf #1} (#2) #3}
\newcommand\PR[3]{{\sl Phys. Reports} {\bf #1} (#2) #3}
\newcommand\AoM[3]{{\sl Ann. of Math.} {\bf #1} (#2) #3}
\newcommand\UMN[3]{{\sl Usp. Mat. Nauk} {\bf #1} (#2) #3}
\newcommand\FAP[3]{{\sl Funkt. Anal. Prilozheniya} {\bf #1} (#2) #3}
\newcommand\FAaIA[3]{{\sl Functional Analysis and Its Application} {\bf #1}
(#2) #3}
\newcommand\BAMS[3]{{\sl Bull. Am. Math. Soc.} {\bf #1} (#2) #3}
\newcommand\TAMS[3]{{\sl Trans. Am. Math. Soc.} {\bf #1} (#2) #3}
\newcommand\InvM[3]{{\sl Invent. Math.} {\bf #1} (#2) #3}
\newcommand\LMP[3]{{\sl Letters in Math. Phys.} {\bf #1} (#2) #3}
\newcommand\IJMPA[3]{{\sl Int. J. Mod. Phys.} {\bf A#1} (#2) #3}
\newcommand\AdM[3]{{\sl Advances in Math.} {\bf #1} (#2) #3}
\newcommand\RMaP[3]{{\sl Reports on Math. Phys.} {\bf #1} (#2) #3}
\newcommand\IJM[3]{{\sl Ill. J. Math.} {\bf #1} (#2) #3}
\newcommand\TMP[3]{{\sl Theor. Mat. Phys.} {\bf #1} (#2) #3}
\newcommand\JPA[3]{{\sl J. Physics} {\bf A#1} (#2) #3}
\newcommand\JSM[3]{{\sl J. Soviet Math.} {\bf #1} (#2) #3}
\newcommand\MPLA[3]{{\sl Mod. Phys. Lett.} {\bf A#1} (#2) #3}
\newcommand\JETP[3]{{\sl Sov. Phys. JETP} {\bf #1} (#2) #3}
\newcommand\JETPL[3]{{\sl  Sov. Phys. JETP Lett.} {\bf #1} (#2) #3}
\newcommand\PHSA[3]{{\sl Physica} {\bf A#1} (#2) #3}
\newcommand\PHSD[3]{{\sl Physica} {\bf D#1} (#2) #3}
\vspace{-1cm}
\noindent
\vskip .3in

\begin{center}

{\large\bf Symmetry Flows, Conservation Laws and }
\end{center}
\begin{center}
{\large\bf Dressing Approach to the Integrable Models}
\end{center}
\normalsize
\vskip .4in

H. Aratyn${}^1$, J.F. Gomes$^{\,2}$, 
E. Nissimov${}^{3}$, S. Pacheva${}^{3}$
and A.H. Zimerman$^{\,2}$
\par \vskip .2in \noindent
${}^1\,\,$Department of Physics, University of Illinois at Chicago,
845 W. Taylor St.\\ 
\hspace*{3mm}Chicago, IL 60607-7059 \\
${}^{2}\,\,$Instituto de F\'\i sica Te\'orica -- 
IFT/UNESP, Rua Pamplona 145 \\
\hspace*{3mm}01405-900, S\~ao Paulo - SP, Brazil\\
${}^{3}\,\,$Institute of Nuclear Research and Nuclear Energy,
Boul. Tsarigradsko Chausee 72\\
\hspace*{3mm}BG-1784 Sofia, Bulgaria\\

\begin{abstract}
The graded affine Lie algebras provide a framework in which the dressing
method is applied to the generic type of integrable models.
The dressing formalism is used to  develop a unified approach
to various symmetry flows encountered among the integrable
hierarchies and to describe related conservation laws.
\end{abstract}

\sect{Introduction}

We consider the integrable hierarchies of differential equations 
in the framework of a linear spectral eigenvalue equation defined 
in terms of the matrix Lax operator.
In order to develop this construction of integrable models, it is, at the 
bare minimum, necessary to define a notion of affine algebra with a graded structure 
together with a semisimple element $E$ of this algebra with a fixed positive grade 
(this grade will be chosen for simplicity to one). 
Such an algebraic approach was coined a generalized 
Drinfeld-Sokolov method in a series of papers \ct{GIH1,GIH2}, which extended 
early results of Drinfeld, Sokolov and Wilson \ct{D-S,wilson81}. 
By including various non-standard gradations, the scheme was shown to 
encompass a class of the constrained KP hierarchies \ct{AFGZ-jmp}. 
The simplicity and solvability of the integrable model in the algebraic 
formalism  ultimately springs from the fact that its building block-the Lax operator- 
can be rotated into much simpler abelian Lax operator by employing the so-called 
dressing transformation. 
Supplementing the algebraic method by the dressing transformation gives 
rise to a simple and elegant construction of all symmetry flows of the
integrable models, including the isospectral deformations of the underlying Lax 
matrices.
The conservation laws follow easily and, via cohomological 
arguments, find a transparent interpretation in terms of the tau-function. 
The scheme involves only the dressing transformation operators in associating
a symmetry flow to every algebra element which commutes with a semisimple 
element $E$. 
For some gradations such algebra elements  
constitute a non-abelian sub-algebra and accordingly the corresponding 
symmetry flows span a non-abelian algebra.
The method extends easily to incorporate the Virasoro symmetry of the 
integrable models. Choosing the underlying algebra 
to be a super-algebra results in the fermionic symmetry flows. 

This purely algebraic approach is shown to reconcile straightforwardly 
with the framework 
of additional symmetries obtained within the calculus of pseudo-differential 
operators \ct{ourvirasoro,AGNP00}. 

The significant portion of our presentation should not come as a surprise to 
experts in the field, see f.i. papers 
\ct{GIH1,GIH2,HMG93,DG98,Miramontes:1999hu,MM99}. 
What we believe has a claim to novelty in this presentation is a unified 
approach to the subject of symmetry flows which naturally includes the 
non-abelian flows and their connection to the tau-functions. 
We also point out a link to an alternative pseudo-differential approach 
to the symmetries of the Lax hierarchies.

\sect{Dressing Technique}
Let $\cgh$ be the affine Lie algebra with an integral gradation:
$ \cgh= \oplus_{n \in \IZ} \, \cgh_n$ with respect to the grading operator
$\qs$ such that $\lb \qs \, , \, \cgh_n \rb = n\, \cgh_n$.
For the background material on
gradation in the framework of the affine Lie algebras see Section [\ref{background}].

We consider the matrix  Lax operator
\be
L= D_x + E + A
\lab{lax-uno}
\ee
which as will be shown below defines an integrable hierarchy associated to 
the linear spectral problem :
\be
L \Psi_0 = \(\pa_x + E + A\) \Psi_0 = 0
\lab{spec}
\ee
For the class of models with $ \cgh = {\widehat {sl}}(M+K+1)$
the corresponding choice of elements $E$ and $A$
is described in Section [\ref{background}] together with other basic
information.
The semisimple element $E$ defines a direct sum decomposition of the 
algebra $\cgh$:
\be
\cgh = \cK \oplus \cM \,,
\lab{decomp}
\ee
where $\cM= \ime$ and $\cK$ is a centralizer of $E$:
\be
\cK = \kere \equiv \{ x \in \cgh \v \sbr{x}{E}= 0 \} \, .
\lab{ckdef}
\ee
{}From Jacobi identities we find the algebraic relations:
\be
\sbr{\cK}{\cK} \subset {\cK} \quad ;\quad \sbr{\cK}{\cM} \subset {\cM}
\lab{subalg}
\ee
For simplicity of presentation we work in this paper with 
$E$ of grade one only ($E \in \cgh_1$).
$\cK$ is a graded sub-algebra of $\cgh$, i.e. 
$ \cK= \oplus_{n \in \IZ} \, \cK_n$.
We will allow here gradations which are more general than the principal
gradation.
For that reason we do not assume that $\cK$ itself is abelian
and therefore in general it differs from its (abelian) center defined
as :
\be
\cC (\cK) \equiv \{ x \in \cK \v \lb x , y \rb = 0 \, , \, 
\forall y \in \cK \} \,. 
\lab{center}
\ee

The potential $A$ in $L$ is chosen to belong to the grade zero component of
$\cM$ ($\cM_0$) in the grade zero component of $\cgh$ ($\cgh_0$).

Based on the above setup one can show that the Lax operator $L$
can be  gauge-rotated into $\kere$ by a dressing transformation
given by ${\rm Ad} (U^{-1})$:
\br
L= D_x + E + A  &\to &L_K = U^{-1} \, L \, U
= D_x + E + K^{-} \lab{rotate} \\
 K^{-} &\equiv &  \sum^{\infty}_{j=1} K^{(-j)} \in \cK_{-}
\lab{kmindef}
\er
where $\cK_{-}= \oplus_{j=-1}^{-\infty} \cK_j$ is a negative part of $\cK$
w.r.t. to the given grading.
Let $U$ be an exponentiation of negative grade generators,
$U = e^{\ug}$, with $\ug =  {\sum_{j=1}^{\infty} u^{(-j)}}$
and $u^{(-j)} \in \cM_{-j}$. 
Note, the absence of the grade zero components in $K^{-}$. This follows from
the fact that the projection of $U^{-1} \, L \, U$ on grade zero is given by
$\sbr{E}{u^{(-1)}} +A$ and lies entirely in $\ime$.
This expression can be put to zero by appropriately choosing
the $\cM$ component $u^{(-1)}$ as a local function of $A$.
{}From identities
\br
\( U^{-1}  E U \)_{-j} &=& \sum_{r=1}^{j+1} (-1)^r {1\o r!} \sum_{k_i :\sum k_i =j+1}
\lb u^{(-k_1)} , \lb u^{(-k_2)} , {\ldots} \lb u^{(-k_r)} , E \rb {\ldots}
\rb \nonu \\ 
\( U^{-1}  \pa_x U \)_{-j} &=& \sum_{r=1}^{j} (-1)^{r-1}{1\o r!} 
\sum_{k_i :\sum k_i =j}
\lb u^{(-k_1)} ,  {\ldots} \lb u^{(-k_{r-1})} , 
\pa_x u^{(-k_r)} \rb {\ldots} \rb \nonu 
\er
we find that the generic expression at the grade $-j$ in \rf{rotate} 
must be of the form 
$\pa_x u^{(-j)} + \sbr{E}{u^{(-j-1)}} +\sbr{A}{u^{(-j)}} + {\ldots} 
= K^{(-j)}$
where omitted terms contain only $u^{(-l)}, l < j$.
Due to the fact that $K^{-} \in \cK$, this recursion procedure allows
the choice of ${u^{(-j-1)} \vert_{\cK}}=0$.
The remaining $\cM$-component $u^{(-j-1)}$ is given in
terms of previously known elements $u^{(-l)}, l \leq j$.
Note, that all elements $u^{(-j)}$ besides belonging to $\cM$,
are local expressions of polynomials of components of $A$.

Next, we proceed to ``gauge away'' the term $K^{-}$ in \rf{rotate}
using that, according to relation \rf{subalg}, $\cK$ is a sub-algebra of $\cgh$.
Following the standard arguments, the dressing of the Lax operator $L_K$
proceeds according to :
\be
S^{-1} \, L_K \, S =  D_x + E
\lab{rotate1}
\ee
where $S= e^{\sg}$ is an exponentiation of negative
grade generators from $\cK$, so that
$\sg = {\sum_{j=1}^{\infty} s^{(-j)}} \in \cK$.
Indeed,  contribution to $S^{-1} \, L_K \, S$ at grade $-1$ is $\pa_x
s^{(-1)} + K^{(-1)}$, which determines $s^{(-1)}$.
At grade $-2$ we find
$\pa_x s^{(-2)} +K^{(-2)}+ \sbr{K^{(-1)}}{s^{(-1)}} + 
\h \sbr{\pa_x s^{(-1)}}{s^{(-1)}}$,
which can be put to zero by the appropriate choice of $s^{(-2)}$.
This process can be continued recursively. For abelian $\cK$ it yields
$K^{-}= -S^{-1} \pa_x S$.
Note, that in contrast to exponential $U$ in \rf{rotate}, the 
exponential $S$ in \rf{rotate1} has a non-local character.

Combining results of eqs.\rf{rotate} and \rf{rotate1} 
we arrive at the solution to the problem :
\be
\Theta^{-1} \,\(D+E+A \) \,\Theta = D+E
\lab{thlth}
\ee
with transformation $\Th \equiv  US$  given by expansion in the
terms of negative grading such that
$ \Th = \exp \( \sum_{l <0} \th^{(l)} \) = 1 + \th^{(-1)} + \ldots $.

Let now $b_n$ be in the center $\cC_n (\cK)$ of $\cK_n$ and $\Theta$ the
dressing operator satisfying eq.\rf{thlth}.
We associate to $\Theta b_n {\Theta}^{-1}$ the following 
descending expansion in grading:
\be 
\Theta b_n {\Theta}^{-1} = b_n + \sum_{k=0}^{\infty} \b_n^{(n-k-1)}
\lab{tbntinv}
\ee
where the term $\b_n^{(n-k-1)}$ has a $(n-k-1)$-grade, i.e. 
$\sbr{Q_{\bf s}}{\b_n^{(n-k-1)}} = (n-k-1)\,\b_n^{(n-k-1)}$
with respect to the grading operator ${Q_{\bf s}}$ (see f.i. 
\rf{grading}).
Next, we define :
\be 
B_n= \( \Theta b_n {\Theta}^{-1} \)_{+} =  
b_n + \sum_{k=0}^{n-1} \b_n^{(n-k-1)}
\lab{bnplus}
\ee
Note, that since $b_n$ commutes with $S$, the right hand side of eq. \rf{tbntinv} can be
rewritten in an explicitly local form $\Theta b_N {\Theta}^{-1} = U  b_N U^{-1}$.
Also, note that $B_1=\( U E U^{-1} \)_{+}=E+A$.

\sect{Symmetry Flows and Conservation Laws}

\subsection{Symmetry Flows and Isospectral Times}

In this Section we show how to associate a symmetry flow $\d_X$ to any constant 
element $X \in \cK=\kere $ of a positive grade.

\begin{definition}
For a constant element
$X_m$, with grade $m > 0$, belonging to $\cK_m$ we 
define a resolvent of $X_m$ as :
\be
X^{\Th}_m \equiv Ad (\Th) X_m = \Th  X_m {\Th}^{-1}
\lab{defresolv}
\ee
\label{definition:resolv}
\end{definition}

Next, apply ${\rm Ad}\, ( \Th)$ on the bracket:
\be
\sbr{X_m}{D_x+E}=0
\lab{xde}
\ee
to obtain:
\be
\lb L \, , \, X^{\Th}_m \rb =0
\lab{resolvent}
\ee

\begin{definition}
Let $X_m$ be in $\cK_m$. Define a transformation $\d_{X_m}$ 
associated to $X_m$ by
\be
\d_{X_m} \Th = (\Th  X_m {\Th}^{-1})_{-} \Th \to 
\d_{X_m} \Th \equiv (X^{\Th}_m)_{-} \Th \quad , \quad 
m \geq 0 
\lab{sym-flows}
\ee
\label{definition:sflows}
\end{definition}

To $b_n$ in the center $\cC_n (\cK)$ of $\cK_n$ we associate 
a flow $\d_{b_n} \equiv d /d t_n$ :
\be
\d_{b_n} \Theta = {d \o d t_n} \Theta = (\Theta b_n {\Theta}^{-1})_{-} \Theta
= \Theta b_n - B_n \Theta 
\lab{bnsym-flows}
\ee
and similarly
\be
{d \o d t_n} {\Theta}^{-1} = - b_n \, {\Theta}^{-1}+  {\Theta}^{-1}\, B_n 
\lab{bnsym-inv}
\ee
Note, that eq. \rf{thlth} is equivalent to $ \pa_x \Theta = 
\Theta E - (E+A) \Theta $
and since $B_1 = E+A$ the definitions \rf{thlth} and \rf{bnsym-flows} imply
that $d / d t_1 = \pa_x$.

Note also, that according to the definition \rf{bnsym-flows}, $B_n$ can be 
rewritten as 
\be
B_n = \Theta b_n {\Theta}^{-1} + \Theta \({d \o d t_n} {\Theta}^{-1}\)
\lab{Bngauge}
\ee

The action of the transformation $\d_{X_m}$ applied
on the potential $A$ is described by the following Lemma.
\begin{lemma} Let $X_m \in \cK_m$, then
\be
\d_{X_m} A = \lb L \, , \, (X^{\Th}_m)_{+} \rb 
\lab{symm-a}
\ee
\label{lemma:symm-a}
\end{lemma}
\begin{proof}
The proof of \rf{symm-a} goes as follows. First from 
eq.\rf{sym-flows} we find
\be
0 = \d_{X_m} \( \Th^{-1} \, L \, \Th\) = - \lb  \Th^{-1} \d_{X_m} \Th  \, , \, \Th^{-1} \, L \,
\Th \rb + \Th^{-1} \, \( \d_{X_m}  L \)\, \Th
\lab{symm-pr1}
\ee
which for $\d_{X_m}  L = \d_{X_m} A$ gives 
\be
\d_{X_m} A = \Th \lb  \Th^{-1} \d_{X_m} \Th  \, , \, \Th^{-1} \, L \,
\Th \rb  \Th^{-1} =  \lb  \d_{X_m} \Th\, \Th^{-1}\, , \, L \rb
\lab{symm-pr2}
\ee
Eq. \rf{symm-a} follows now by virtue of the resolvent identity
\rf{resolvent} and definition (\ref{definition:sflows}).
\end{proof}

Consider, again a constant element
$b_{n}$, of grade $n$ ($n>0$) such that $b_{n} \in \cC_n (\cK)$.
{}From the above considerations one gets 
\be
{d A \over d t_n }= \lb L \, , \, B_n \rb
\lab{iso}
\ee
The commutative symmetry transformations in \rf{bnsym-flows} and \rf{iso} are called 
{\it the isospectral flows}.
Note, that the identity $ U \, b_{N}\, U^{-1} =   b_{N}^{\Th}$ ensures
locality of the corresponding conservation laws.

It is natural to generalize the dressing relation \rf{thlth} to other
isopectral flows by defining 
\be
L_N \equiv \Th \( {d \over d t_N} +  b_N \) {\Th}^{-1}
=  {d \over d t_N} -  \( {d \over d t_N}  \Th \)  {\Th}^{-1}
+ U b_N U^{-1}
\lab{mndef}
\ee
which coincides with \rf{thlth} for $N=1$.
{}From \rf{Bngauge} we find 
\be
L_N =  {d \over d t_N} + B_N
= \Psi_0 {d \over d t_N}\Psi^{-1}_0
\lab{mnexpr}
\ee
for the wave-function $\Psi_0$ :
\be
\Psi_0 \equiv  \Th \exp \( - \sum_{N=1}^{\infty} b_N t_N \)
\qquad b_1 = E  \quad ; \quad t_1 =x
\lab{BA-function}
\ee
defined in terms of the parameters
$t_N$ such that $\sbr{d/d t_{N^{\pr}}}{t_N} = \d_{N^{\pr} N}$.
Due to \rf{mndef} such wave-function $\Psi_0$ satisfies :
\be 
L_N \Psi_0 =0  \qquad N=1,2,{\ldots} 
\lab{mnpsi}
\ee
and therefore is a solution for the underlying linear spectral
problem \rf{spec}.
Eqs.\rf{mnpsi} are equivalent to an hierarchy of evolution equations :
\be
{d \Psi_0 \over d  t_N } = - B_N \Psi_0
\lab{hierarchy}
\ee
Moreover, by conjugating with ${\rm Ad} (\Th)$ identities:
\br 
\sbr{{d \over d t_N} +  b_N }{D_x+E}  &=& 0 \lab{tnbnxe}\\
\sbr{{d \over d t_N} +  b_N }{{d \o d {t_M}} +  b_M} &=& 0  \lab{tnbntmbm}
\er
we obtain the Zakharov-Shabat equations
\br
\sbr{L_N}{L} \!\! &=& \!\! 
{d {A} \o d {t_N}} -\pa_x B_N +\sbr{B_N }{E+A} =0 \lab{xtn}\\
\sbr{L_N}{L_M} \!\! &=& \!\! 
{ d {B_M} \o d {t_N} }- {d {B_N} \o d {t_M}} +\sbr{B_N }{B_M} =0 \lab{tntm}
\er
We recognize in these equations compatibility conditions for the linear relations \rf{spec} and 
\rf{mnpsi}.

We will now use \rf{symm-a} to show that eq. \rf{sym-flows} for
arbitrary $X_m \in \cK$ generates a 
well-defined symmetry transformation of the above model, meaning that
\begin{itemize}
\item $\d_{X_m} A \in \cM_0$ 
\item  transformations $\d_{X_m}$ commute
with the isospectral flows.
\item  transformations $\d_{X_m}$ close into an algebra. 
\end{itemize}
In other words we have the following Lemma
\begin{lemma}
The transformations in \rf{sym-flows}  (or in \rf{symm-a}) are symmetry
transformations of the model defined by $L \Psi_0 =0$
and $ d L /d t_N = \lb L \, , \, B_N \rb$. 
\label{lemma:sym-model}
\end{lemma}
\begin{proof}  
The proof goes as follows. First notice that \rf{resolvent} implies:
\be
\lb L \, , \, (X^{\Th}_m)_{+} \rb = - \lb L \, , \, (X^{\Th}_m)_{-} \rb
\lab{resolvent-1}
\ee
The conventional ``dressing'' argument compares
grades on both sides of equation \rf{resolvent-1}.
The left hand side involves terms with grades $ \geq 0$ 
while grades on the right hand side are between $0$ and $- \infty$.
Consequently, each side of eq.\rf{resolvent-1} lies in the zero grade sub-algebra.
Correspondingly, the contributions of the above terms are equal to:
\be
\d_{X_m} A = \lb D_x +A \, , \, (X^{\Th}_m)_{0} \rb = - \lb E \, , \, (X^{\Th}_m)_{-1} \rb 
\lab{resolvent-2}
\ee
The last equality ensures that $\d_{X_m} A$ is in $\cM_0$ and therefore
the transformation generated by \rf{symm-a} or \rf{symm-pr2}
is well-defined.

To complete the proof of Lemma (\ref{lemma:sym-model})
we will show that the algebra of transformations from \rf{sym-flows}
closes and commutes with the isospectral flows.
Let us first discuss the algebra closure.
Consider:
\be
\( \d_{X_m} \d_{X_n} - \d_{X_n} \d_{X_m} \) \Th = 
\d_{X_m} \( (X^{\Th}_n)_{-} \Th \) - \d_{X_n} \( (X^{\Th}_m)_{-} \Th \)
\lab{sym-closure}
\ee
To proceed we need to find  $\d_{X_n} (X^{\Th}_m)_{-} $.
\br
\d_{X_n} \( \Th X_m \Th^{-1} \)_{-} &=& \lb  (\d_{X_n} \Th) \Th^{-1}  \, , \, \Th \,X_m \,
\Th^{-1} \rb_{-} =
\lb  (X^{\Th}_n)_{-}  \, , \,  X^{\Th}_m  \rb_{-} \nonu \\
&=& \lb  (X^{\Th}_n)_{-}  \, , \,  (X^{\Th}_m)_{-}  \rb +
\lb  (X^{\Th}_n)_{-}  \, , \,  (X^{\Th}_m)_{+}  \rb_{-} 
\er
Inserting these results into \rf{sym-closure} we get
\be
\( \d_{X_m} \d_{X_n} - \d_{X_n} \d_{X_m} \) \Th = f_{mn}^{k} (X^{\Th}_k)_{-} \Th
= f_{mn}^{k} \d_{X_k} \Th = \d_{\sbr{X_m}{X_n}} \Th
\lab{dmdndk}
\ee
after comparing with the algebra of generators in $\cK$ 
\be
\sbr{X_m}{X_n} = f_{mn}^{k} X_k
\lab{xmxnxk}
\ee
and noticing that the left hand side after dressing by $\Th$ and projecting on
the negative modes becomes :
$$
{\sbr{X^{\Th}_m}{X^{\Th}_n}}_{-} = \lb  (X^{\Th}_m)_{-}  \, , \,  (X^{\Th}_n)_{-}  \rb
+\lb  (X^{\Th}_m)_{-}  \, , \,  (X^{\Th}_n)_{+}  \rb_{-}+
\lb  (X^{\Th}_m)_{+}  \, , \,  (X^{\Th}_n)_{-}  \rb_{-}
$$

Now, consider commutation with the isospectral flows. Since $b_N \in \cC (\cK)$
we have 
\be
\sbr{X_m}{b_N} = 0
\lab{xmbn}
\ee
and the same arguments as above yield this time:
\be
\( \d_{X_m} {d \o d t_N} - {d \o d t_N} \d_{X_m} \) \Th = 0
\lab{com-iso}
\ee
\end{proof}
\subsection{Conservation Laws}
We now associate to each  $X_n \in \cK$ the following class of objects.
\begin{definition}
Define maps $\cJ$, $\O$: $\cK \to \IC$ as :
\br
\cJ (X_n)\, &\equiv& \,  \Tr \( \sbr{Q_{\bf s}}{\Theta}\, X_{n} {\Theta}^{-1} \) 
\lab{cjxn}\\
\O (X_n) \, &\equiv& \, -  \Tr \( E X^{\Th}_n \) \, = \, -  \Tr \( E {\Th} X_n {\Th}^{-1} \) 
\lab{omn} 
\er
\label{definition:omegas}
\end{definition}
Here, $\Tr ( {\ldots} )= \tr ({\ldots} )_0$ involves both the conventional
matrix trace operation $\tr$ as well as a projection on the zero grade.

The above objects  are related through :
\begin{proposition}
\be
\pa_x \cJ (X_n) = \O (X_n) \lab{claim}
\ee
\label{proposition:pxjnnh}
\end{proposition}

\begin{proof}
The proof follows by taking $m=1$ in eqs. \rf{bnsym-flows}
and \rf{bnsym-inv} which produces:
\be
\pa_x \cJ (X_n)  = - \Tr \( \sbr{Q_{\bf s}}{ B_1 }\, \Theta \, X_{n}\, {\Theta}^{-1} \) 
\lab{paxjna}
\ee
Since $B_1$ is equal to $E + A$ and $\sbr{Q_{\bf s}}{E+A}=E$
we get from eq. \rf{paxjna}.
$\pa_x \cJ (X_n)  = -\Tr \( E \Theta \, X_n\,  {\Theta}^{-1}\, \) = \O (X_n)$
\end{proof}
\begin{proposition}
\be
\d_{X_n} \cJ (X_m)   - \d_{X_m} \cJ (X_n) = f_{nm}^{k} \cJ (X_k)
\lab{jno}
\ee
where $ f_{mn}^{k}$ is the structure constant of the sub-algebra $\cK$ 
from relation \rf{xmxnxk}.
\label{proposition:dnjm-dmjn}
\end{proposition}
\begin{proof}
We make use of the property:
$ \Tr \( A \, \sbr{Q_{\bf s}}{ B} \) = - \Tr \(  \sbr{Q_{\bf s}}{ A}\, B \)$,
satisfied by the trace $\Tr$.
Accordingly,
\be
\d_{X_n} \cJ (X_m) 
 =- \Tr \(\(\Theta \, X_{n}\,{\Theta}^{-1}\)_{-} \sbr{Q_{\bf s}}{ \Theta
\, X_{m}\, {\Theta}^{-1}} \)
\lab{dxnjxm}
\ee
and therefore 
\be
\d_{X_n} \cJ (X_m)   - \d_{X_m} \cJ (X_n) =  \Tr \(\(\Theta \, X_{m}\,{\Theta}^{-1}\) 
\sbr{Q_{\bf s}}{ \(\Theta \, X_{n}\,{\Theta}^{-1}\)} \) 
\lab{dnjmdmjn}
\ee
which is equal to
\br 
&&\Tr \(\(\Theta \, X_{m}\,{\Theta}^{-1}\) 
\sbr{Q_{\bf s}}{ \Theta} \, X_{n}\,{\Theta}^{-1} \) 
+\Tr \(\(\Theta \, X_{m}\,{\Theta}^{-1}\) 
 \Theta \, n X_{n}\,{\Theta}^{-1} \) \nonu \\
&- &\Tr \(\(\Theta \, X_{m}\,{\Theta}^{-1}\) 
{ \Theta} \, X_{n}\,{\Theta}^{-1} \sbr{Q_{\bf s}}{ \Theta} {\Theta}^{-1}\) 
\lab{dnjmdmjna}
\er
The middle term  vanishes being equal to $n \Tr ( X_m X_n)$
while the first and last terms combine to give :
\be
\Tr \( \sbr{Q_{\bf s}}{\Theta}\, \( X_{n}X_{m}-X_{m}X_{n}\) {\Theta}^{-1} \) 
= \cJ \( \sbr{X_{n}}{X_{m}} \)
\lab{jxnxm}
\ee
as announced in the proposition.
\end{proof}

Result of the Proposition \ref{proposition:dnjm-dmjn} can be reformulated in the
following more formal statement.
\begin{lemma}
A one-form $\cJ (\cdot)$ on $\cK$ with values in $\IC$ defined in 
(\ref{definition:omegas}) is a closed one-cocycle:
\be
d \cJ (X_n, X_m ) = 0
\lab{dcjxx}
\ee
with respect to the usual Cartan-Chevalley-Eilenberg differential $d$ given 
by the formula:
\br
d \cJ (X_1, {\ldots} , X_n) &=& \sum_{i=1}^n (-1)^{(i-1)} \d_{X_i} 
\cJ (X_1, {\ldots} , {\hat X_i}, {\ldots} , X_n) \lab{dcjxxa} \\
&+&
\sum_{j <k} (-1)^{(j+k)} \cJ \(  \lb  X_j  \, , \,  X_k  \rb, {\ldots} ,
{\hat X_j}, {\ldots} ,{\hat X_k}, {\ldots} ,X_n \)
\nonu 
\er
\label{lemma:chevalley}
\end{lemma}
As a corollary of Proposition \ref{proposition:pxjnnh} we obtain
that $\O (X_n)$ too is a closed one-cocycle 
\be
d \O (X_n, X_m ) = 0 \,.
\lab{doxx}
\ee
We will now address a question whether the closed cocycle $\O (X_n)$
is also exact, namely whether there exists $\O_0$ such that
\be
 \O (X_n) = {d } \, \O_0  (X_n)
\lab{dnoxn}
\ee
{}From the definition (\ref{definition:sflows}) it follows that
$\d_{X_m} \Th = (US  X_m {S}^{-1} U^{-1})_{-} US$.
Assuming that $\d_{X_m}$ acts as a derivative (satisfies Leibniz rule) we get:
\be
{S}^{-1} \d_{X_m} S = - {S}^{-1}( U^{-1} \d_{X_m} U) S+ {S}^{-1} U^{-1}
(US  X_m {S}^{-1} U^{-1})_{-} US
\lab{leib}
\ee
After multiplying by $E$ and taking the trace eq. \rf{leib} becomes:
\be
\Tr \( E \, {S}^{-1} \d_{X_m} S \) = -\Tr \( E \, U^{-1} \d_{X_m} U\)+ \Tr \( E \,
 U^{-1} (US  X_m {S}^{-1} U^{-1})_{-}  U \)
\lab{saway}
\ee
after use was made of cyclicity of the trace and $S E {S}^{-1}= E$.

Let us now consider the second term on the right hand side of 
\rf{saway}.
Let $ \cP_{(-1)}$ be  a projection on grade $-1$:  $ \cP_{(-1)}(O) = O_{-1}$.
It is clear that $\cP_{(-1)} \(U^{-1} \(U  S  X_m {S}^{-1} U^{-1}\)_{-} U \)
= \cP_{(-1)} \( \(U  S  X_m {S}^{-1} U^{-1}\)_{-} \)$.
We can therefore rewrite \rf{saway} as:
\be
\Tr \( E \, {S}^{-1} \d_{X_m} S \) = -\Tr \( E \, U^{-1} \d_{X_m} U\)+ \Tr \( E \,
(US  X_m {S}^{-1} U^{-1})_{-}  \)
\lab{sawayb}
\ee
The only non-zero contribution from the first term is
$ \Tr \(E  \d_{X_m} u^{(-1)} \)$ since   $\d_{X_m} u^{(-1)}$ is the
only term in $U^{-1} \d_{X_m} U$ of grade $-1$.
However, since $ u^{(-1)} \in \cM$ the trace product of $E$ with
$\d_{X_m} u^{(-1)}$ yields zero. Hence, there is no contribution from the first term
on the right hand side of equation \rf{sawayb}.
Similarly, the remaining term on the right hand-side 
of eq. \rf{sawayb} is
$ \Tr \(E  \d_{X_m} \sg^{(-1)} \) = \d_{X_m} \Tr \(E  \sg^{(-1)} \) $.
Hence, \rf{sawayb} is equivalent to:
\be
\d_{X_m} \Tr \( E \, \sg \) =  \Tr \( E \, (\Th  X_m \Th^{-1} )_{-}  \)
\lab{sawayc}
\ee
which is nothing but the statement that $\O (\cdot)$ is exact and reproduced by
\be
\O (X_m) = - \d_{X_m} \Tr \( E \, \sg \)= - \d_{X_m} \Tr \( E \, \sg^{(-1)} \)
\lab{exact}
\ee
\subsection{Isospectral Flows and Conservation Laws}
For the special case of $X_n = b_n \in \cC (\cK)$ definition 
\ref{definition:omegas} becomes :
\br
\cJ_n &\, \equiv \,&  \Tr \( \sbr{Q_{\bf s}}{\Theta}\, b_{n} {\Theta}^{-1} \) 
\lab{jns-general}\\
\cH_n &\, \equiv \,&  \O (b_n) =-  \Tr \( E U b_n U^{-1}  \) \lab{hams} 
\er
Note, that $\cJ_n$ depends explicitly on $S$ and is therefore in general  a non-local quantity
in contrast to $\cH_n$.

{}From Proposition \ref{proposition:pxjnnh} we obtain :
\be
\pa_x \cJ_n= \cH_n \lab{claima}
\ee

\begin{proposition}
\be
{d \o d t_m} \cJ_n = - \Tr \( \sbr{Q_{\bf s}}{ B_m }\, \Theta \, b_{n}\, {\Theta}^{-1} \) 
\lab{pamjn-general}
\ee
\end{proposition}
\begin{proof}
The proof follows from the direct calculation:
\br
{d \o d t_m} \cJ_n &= & \Tr \( \sbr{Q_{\bf s}}{(\Theta b_m - B_m \Theta )}\, b_{n}\, 
{\Theta}^{-1} \) \nonu\\
&+&\Tr \( \sbr{Q_{\bf s}}{\Theta}\, b_{n}\, 
(- b_m \, {\Theta}^{-1}+  {\Theta}^{-1}\, B_m \) 
\lab{pamjn}
\er
where we used relations \rf{bnsym-flows}-\rf{bnsym-inv}.
Eq. \rf{pamjn-general} follows now from commutativity of $b_n$ and $b_m$
and also due to the fact that $\Tr (n b_n b_m)=0$.
\end{proof}

Note, that expression in equation \rf{pamjn-general} can be rewritten as
\be
{d \o d t_m} \cJ_n = - \Tr \( \sbr{Q_{\bf s}}{ B_m }\, U \, b_{n}\, U^{-1} \) 
\lab{pamjn-generalu}
\ee
which clearly exhibits the local character of $ d \cJ_n / d t_m$.

{}From Proposition \ref{proposition:dnjm-dmjn} applied to the abelian center
of $\cK$ follows a set of corollaries :
\begin{corollary}
\br
{d \o d t_m} \cJ_n &= &{d \o d t_n} \cJ_m \lab{pamjn=panjm} \\
{d \o d t_m} \cH_n  &= & {d \o d t_n} \cH_m \lab{commlaws}
\er
\label{corollary:pamjn=panjm}
\end{corollary}
Inserting expansions \rf{tbntinv} and \rf{bnplus}
into relation \rf{pamjn-general} we obtain
\br
{d \o d t_m} \cJ_n &=& - m \tr \( b_m \b_n^{(-m)} \) -\sum_{k=0}^{m-1} 
(m-k-1) \tr \(  \b_m^{(m-k-1)} \b_n^{(k+1-m)} \) 
\nonu \\ 
{d \o d t_n} \cJ_m &=& - n \tr \( b_n \b_m^{(-n)} \) -\sum_{k=0}^{n-1} 
(n-k-1) \tr \(  \b_n^{(n-k-1)} \b_m^{(k+1-n)} \) 
\nonu  
\er
Taking $m=1$ in the above equations and equating them according to
relation \rf{pamjn=panjm} we get the recurrence relation:
\be
\tr \(E\b_n^{(-1}\) =  n \tr \( b_n \b_1^{(-n)} \) -\sum_{k=0}^{n-1} 
(n-k-1) \tr \(  \b_n^{(n-k-1)} \b_1^{(k+1-n)} \) 
\lab{pancjmpamcjn}
\ee

A conservation law has the form
\be
{d \o d t_m} \cH_n + \pa_x Q_{m,n} =0 
\lab{conservat}
\ee
with local, conserved flux $Q_{m,n}$ and conserved (Hamiltonian)
density $\cH_n $.
Relations \rf{claima} and \rf{pamjn-generalu} establish
existence of an infinite number of local conservation laws
for all integrable models obtained by the algebraic dressing construction.
The local, conserved flux is given by 
$Q_{m,n}= \Tr \( \sbr{Q_{\bf s}}{ B_m }\, U \, b_{n}\, U^{-1} \) $.
The following corollary follows easily.
\begin{corollary}
The Hamiltonians defined by
\be
H_n = \int \cH_n \; dx \quad ; \quad n=1,2, {\ldots} 
\lab{conslaws}
\ee
are conserved.
\end{corollary}
\begin{proof}
Indeed
\be
{d \o d t_m} H_n = \int \pa_x {d \o d t_m} \cJ_n
\lab{dtmhn}
\ee
The desired result follows now recalling that 
the quantities $ d \cJ_n / d t_m$ are local as shown in \rf{pamjn-generalu}.
\end{proof}

Let us apply $\cH_n= - \Tr \(E  U  b_n U^{-1} \)$ to the special case of $n=1$: 
\br
\cH_1&=& - \Tr \(E  U  E U^{-1} \) = \Tr \(E  \sbr{E}{u^{(-2)}} \)
\nonu \\
&-& \h \Tr \(E \sbr{u^{(-1)}}{\sbr{u^{(-1)}}{E}} \)
\lab{hamnone}
\er
By well-known trace identity the first term is zero and the second term
becomes
\be
\cH_1 = \h \Tr \({\sbr{u^{(-1)}}{E}}^2 \) = \h  \Tr \( A^2 \)
\lab{hamnonea}
\ee
which is valid for models described by the Lax operator $L$
from \rf{rotate}.

One important consequence of Corollary \ref{corollary:pamjn=panjm} is
that $\cJ_n$ appears to be ${d \o d t_n}$ of some function of phase variables. 
The standard way of writing this is in terms of the tau-function:
\begin{definition}
\be    
\cJ_n= - {d \o d t_n} \log \tau  \,.
\lab{jn-tau}
\ee
\label{definition:tau-def}
\end{definition}
We therefore have:
\be
\Tr \( U^{-1} \sbr{Q_{\bf s}}{U}\, b_{n}  \) 
+ \Tr \( S^{-1} \sbr{Q_{\bf s}}{S}\, b_{n}  \) 
= - {d \o d t_n} \log \tau
\lab{ustau}
\ee
Recall, that $U=\exp ( \ug)\, ,\, S=\exp ( \sg)$ and
\be
\( d e^{\sg}\) e^{- \sg} = \sum_{n=1}^{\infty} {1 \o n!}
\( {\rm ad}\, \sg \)^{n-1} d \sg 
\lab{helga}
\ee
for a derivation $d$.
Using this one can establish  that the contribution on the
left hand side of \rf{ustau} from the second term is equal
to $\Tr \( \sbr{\qs}{\sg} b_n \)$.
For $n=1,2$ the contribution on the
left hand side of \rf{ustau} from the first term is zero
and accordingly
\br
\tr \( \sg^{(-1)} E\) &= &  \pa_x \log \tau \lab{paxltau}\\
\tr \( \sg^{(-2)} b_2\) &= &  \h {d \o d t_2} \log \tau \lab{patwoltau}
\er
One important consequence of \rf{claim} is that 
\be 
\cH_n =- {d \o d t_n} \pa_x \log \tau = - {d \o d t_n} \tr \( \sg^{(-1)} E\) 
\lab{chnsg}
\ee
where in the last equality we used eq.\rf{paxltau}.
This is in perfect agreement with eq.\rf{exact}
obtained in different way in a general case.

\subsubsection{Homogeneous Gradation}
We now turn to the homogeneous gradation with $E= \l E^{(0)}$, 
$b_n \equiv E^{(n)}=\l^n E^{(0)}$.
In this setup we are working with expansions :
\br
\Theta &= &1 + \sum_{k=1}^{\infty} \theta^{(-k)} / \l^k
\lab{thetexp}\\
\Theta E {\Theta}^{-1} &=& E +A +\sum_{k=1}^{\infty} A^{(-k)} / \l^k
\lab{theth}\\
\Theta b_n {\Theta}^{-1} &=& \l^{n-1} E + \l^{n-1} A +\sum_{k=1}^{\infty}
\l^{n-k-1} A^{(-k)} \lab{thenth}
\er
It follows that:
\be
B_n = b_n + \l^{n-1} A +\sum_{k=1}^{n-1} \l^{n-k-1} A^{(-k)} \lab{bnpl}
\ee
Comparing with expansion \rf{tbntinv} we see that 
$\b_n^{(j)} = \l^j A^{j+1-n}$.

In case of homogeneous gradation the definition \rf{jns-general} becomes :
\be
\cJ_n=  \Tr \( {d  \Theta \o d \l} b_{n+1} {\Theta}^{-1} \) 
\lab{jns}
\ee
while \rf{pamjn-general} simplifies to :
\be
{d \o d t_m} \cJ_n = - \Tr \( {d  B_m \o d \l}\, \Theta \, b_{n+1}\, {\Theta}^{-1} \) 
\lab{panjm}
\ee
Expanding the right hand side of \rf{panjm} we obtain (with $A^{(0)}=A$) :
\be
{d \o d t_m} \cJ_n = - m \tr \( E^{(0)} A^{(1-n-m)} \) -\sum_{k=0}^{m-1} 
(m-k-1) \tr \(  A^{(-k)} A^{(2+k-n-m)} \) 
\lab{panjmexp}
\ee
which is equal to :
$$ 
{d \o d t_n} \cJ_m = - n \tr \( E^{(0)} A^{(1-n-m)} \) -\sum_{k=0}^{n-1}
(n-k-1) \tr \(  A^{(-k)} A^{(2+k-n-m)} \)
$$ 

Another crucial observation is that :
\be
A^{(-n)} = {d \, \theta^{(-1)} \o d t_n} 
\lab{crucial}
\ee
The proof follows by projecting eq.\rf{bnsym-flows} on the $-1$
grade. This gives according to \rf{thenth} :
\be
{d \o d t_n} \theta^{(-1)} \l^{-1}  = \(\Theta b_n {\Theta}^{-1} \)_{-1}
= A^{(-n)}  \l^{-1}
\lab{crucial-proof}
\ee
which leads to the desired relation.

One consequence of \rf{crucial} is a relation
\be
B_n = b_n + \l^{n-1} A +\sum_{k=1}^{n-1} \l^{n-k-1} {d\, \theta^{(-1)} \o d t_k} 
\lab{bnpla}
\ee
from which follows the recurrence relation:
\be 
B_{n+1} = \l B_n + {d \, \theta^{(-1)} \o d t_n} = \l B_n +A^{(-n)}
\lab{bnbn}
\ee

Also
\be
\pa_x \cJ_n = \cH_n =  
- \tr \( E^{(0)} A^{(-n)}\)= - \tr \( E^{(0)} {d \theta^{(-1)} \o d t_n}  \)  
\lab{hntnt}
\ee

Note, that the Hamiltonian densities in the the homogeneous gradation can be 
rewritten as :
\be
\cH_n= - \Tr \(E  U  E^{(n)} U^{-1} \) =
- \Tr \(E^{(0)} U  E^{(n+1)} U^{-1} \) = - \Tr \(E^{(0)} B_{n+1} \)
\lab{wils}
\ee
c.f. \ct{wilson81,AGZ95,ADR97}.
Consider the matrix $A$ as in \rf{a-homo}. Then the general expression
for $\cH_1$ found in \rf{hamnonea} becomes in this case
$\cH_1 = \sum_i^M q_i r_i $.
For the case of $n=2$ ($\cH_2$) we obtain :
$$ 
\cH_2 =  \Tr \( \lb u^{(-2)} , E^{(2)}\rb \lb u^{(-1)}, E  \rb \)
=  \sum_{i=1}^M  \( q_i r_{i \, , \, x} - q_{i \, , \, x} r_i \)
$$ 
which is consistent with $\pa_x \cH_2 = d \cH_1 / d t_2$.

Plugging  $\b_n^{(j)} = \l^j A^{(j+1-n)}$ into the recurrence relation
\rf{pancjmpamcjn} we obtain
$$ 
\tr \( E^{(0)}A^{(-n)} \) =
n \tr \(  E^{(0)} A^{(-n)} \) +\sum_{k=0}^{n-1}
(n-k-1) \tr \(  A^{(-k)} A^{(1+k-n)} \)
$$ 
This recurrence relation can equivalently be written as :
\be
\tr \( E^{(0)}A^{(-n)} \) =- \tr \(  A A^{(1-n)} \)
- \h \sum_{k=1}^{n-2} \tr \(  A^{(-k)} A^{(1+k-n)} \) \;\; \;{\rm for}\,\;
n>1
\lab{recura}
\ee
\subsection{Recursion Relations}
{}From \rf{bnsym-flows} we find :
\be
\d_{b_N} \(\Theta b_M {\Theta}^{-1}\) = - \sbr{B_N}{\(\Theta b_M {\Theta}^{-1}\)}
\lab{dnbm}
\ee
Of special interest is $b_1=E$ and the corresponding conjugated
element $\(\Theta b_N {\Theta}^{-1}\)$, which is given by expansion in
grading (see eq.\rf{tbntinv}):
\br
\Theta b_1 {\Theta}^{-1} = U  E {U}^{-1} &=& E + \sbr{u^{(-1)}}{E} +\sum_{k=1}^{\infty}
\b^{(-k)}
\lab{b1u}\\
 &=&E + A +\sum_{k=1}^{\infty} \b^{(-k)} \nonu
\er
where in the last equation we used that by construction $\sbr{u^{(-1)}}{E}=A$
and for brevity we wrote $\b_1^{(-k)} =\b^{(-k)}$.
Plugging \rf{b1u} into \rf{dnbm} we obtain by projecting on 
grade zero :
\be
\d_{b_N} A = - \sbr{b_N}{\b^{(-N)}}
\lab{dna}
\ee
Hence, only $\cM$ components of $\b^{(-N)}$ will make a non-zero
contribution to the flows of $A$.
For $N=1$ we find from eq. \rf{dna}:
\be
\d_1 A = \pa_x A= - \sbr{E}{\b^{(-1)}}
\lab{donea}
\ee
while from eq. \rf{dnbm} we get
\be
\d_1 \b^{(-n)} = \pa_x \b^{(-n)} = - \sbr{E}{\b^{-(n+1)}} 
- \sbr{A}{\b^{(-n)}}
\lab{doneeun}
\ee
Introducing, the covariant derivative $\cD = \pa_x + ad_{A}$ we can rewrite
\rf{doneeun} in a compact form as:
\be
\cD \b^{(-n)} + ad_{E} \(\b^{-(n+1)}\) = 0
\lab{zsa-sc}
\ee
which decomposes on $\cM$ and $\cK$ directions
(with $\b^{(-n)} = \b_{\cM}^{(-n)} +\b_{ \cK}^{(-n)}$)
as follows:
\be
\b_{ \cM}^{-(n+1)}= - ad^{-1}_{E}  \( (\cD \b^{(-n)}) \v_{\cM}\) \quad ;\quad
(\cD \b^{(-n)}) \v_{\cK} =0
\lab{zsa-scc}
\ee
The first of expressions in \rf{zsa-scc}
can be rewritten as:
\br
\b_{ \cM}^{(-n-1)}  &=&- ad^{-1}_{E}  \( \pa_x \b_{ \cM}^{(-n)} +
\lb A \, , \, \b_{ \cK}^{(-n)} \rb + \lb A \, , \, \b_{ \cM}^{(-n)}\rb
\v_{\cM} \) \lab{zsa-ccc} \\
&=& - ad^{-1}_{E}  \( \pa_x \b_{ \cM}^{(-n)} -
\lb A \, , \, \pa_x^{-1} \( \lb A \, , \,\b_{ \cM}^{(-n)} \rb\v_{\cK} \)
\rb + \lb A \, , \, \b_{ \cM}^{(-n)} \rb
\v_{\cM} \)
\nonu
\er
where we substituted $\b_{ \cK}^{(-n)}$ by:
\be
\b_{ \cK}^{(-n)} = -
\pa_x^{-1} \( \lb A \, , \,\b_{ \cM}^{(-n)}\rb\v_{\cK} \) 
\lab{bmk}
\ee
derived from the second equation in \rf{zsa-scc}.

Since $ \lb A \, , \,\b_{ \cM}^{(-n)}\rb\v_{\cK} 
= \( \cD \b_{ \cM}^{(-n)}  \) \v_{\cK}$
we can rewrite \rf{zsa-ccc} as
\be
\b_{ \cM}^{(-n-1)}  = \cR \( \b_{ \cM}^{(-n)}  \)
\lab{zsa-rec}
\ee
with help of the recursion operator:
\be
\cR = - ad^{-1}_{E}  \( \Pi_{\cM} \cD - ad_{A} \pa_x^{-1} \Pi_{\cK} \cD \)
\lab{rec-1}
\ee
where $\Pi_{\cM}, \Pi_{\cK}$ are projections on $\cM$ and $\cK$
spaces.
This construction simplifies significantly 
when considered in case of the homogeneous gradation (and
symmetric spaces with $\sbr{\cM}{\cM} \subset \cK$).
{}From now on we consider therefore the special case of homogeneous gradation
In this case we have expansion in \rf{bnpl}.
Recall, that for symmetric spaces $ad_{E}^2 = \l^2 I$ on $\cM$.
By applying $ad_E$ on both sides of eq. \rf{doneeun} we
obtain:
\be
A^{(-n-1)} \v_{\cM} = -  \sbr{E^{(0)}}{\pa_x A^{(-n)}+\sbr{A}{A^{(-n)}}}
= - ad_{E^{(0)}} (\cD A^{(-n)})
\lab{covder}
\ee
where $E^{(0)}= \l^{-1} E$.
For the $\cK$ component we get from \rf{doneeun} a non-local expression:
\be
A^{(-n-1)} \v_{\cK} = - \pa_x^{-1} \sbr{A}{A^{(-n)} \v_{\cM}}
\lab{km}
\ee
From above equations obtain the recurrence relation:
\be
A^{(-n-1)} \v_{\cM} = \cR (A^{(-n)} \v_{\cM}) 
\lab{rec}
\ee
with help of the recursion operator:
\be
\cR = - ad^{-1}_{E}  \( \pa - ad_{A} \pa_x^{-1}  ad_{A} \)
\lab{rec-symmetric}
\ee
which is a specialization of $\cR$ in eq. \rf{rec-1} in case of symmetric
spaces.
Also, from \rf{donea} we get :
\be 
A^{(-1)} \v_{\cM} = - ad_{E^{(0)}} (\pa_x A) = \cR ( A))
\lab{bonena}
\ee
and therefore:
\be
A^{(-n)} \v_{\cM} = \cR^n ( A))
\lab{recur}
\ee
which is a well-known recursion relation.

\subsection{Example: AKNS Hierarchy; The Homogeneous Hierarchy with  ${\bf
{\hat sl} (2)=A^{(1)}_1}$}

We take $\cG = sl (2, \IC)$ with standard basis $e = \sigma_{+}$, $f=
\sigma_{-}$ and $h = \sigma_{3}$.
The operator $L=D+E+A$ reads:
\be
L = \twomat{D+\l/2}{q}{r}{D-\l/2} = I \cdot D + {\l \o 2} h + q e +r f
\lab{lopsl2}
\ee
The matrix $U  = \exp \( \sum_{j \geq 1} u^{(-j)}\l^{-j}\)$ with 
\be
u^{(-1)} = \twomat{0}{-q}{r}{0}
\; ;\;
u^{(-2)} = \twomat{0}{q_x}{r_x}{0}\; ;\; {\ldots} 
\lab{uthree}
\ee
transforms $L$ as follows :
\be
U^{-1} L U = \twomat{D+\l/2}{0}{0}{D-\l/2}+ \sum_{i=1}^{\infty} k^{(-i)} \l^{-i} \sigma_3 
\lab{ulusl2}
\ee
where to lowest orders in $\l^{-1}$ we find:
\br
\sum_{i=1}^{\infty} k^{(-i)} \l^{-i} \sigma_3 \!\!&=&\!\! 
\twomat{qr}{0}{0}{-qr}\, \l^{-1} \lab{kmat} \\
&+& \h \twomat{-q_x
r+qr_x}{0}{0}{-qr_x+rq_x}\, \l^{-2} + O ( \l^{-3})  \nonu
\er
We obtain the following expression for $B_2$:
\be
B_2 = (Ub_2U^{-1})_{+} =  \twomat{\l^2/2 - qr}{\l q-q_x}{\l r+r_x}{-\l^2/2 + qr}
\lab{b2sl2}
\ee
The corresponding flows :
\be
\pa_2 q = -q_{xx}+2 q^2r \;\;\; ; \;\;\; \pa_2 r = r_{xx}-2 qr^2 
\lab{secflow}
\ee
reproduce the well-known Nonlinear Schr\"{o}dinger (NLS) equation.

\subsubsection{Tau Functions from the Squared Eigenfunction Potentials}

Let standard AKNS pseudo-differential Lax operator be
\be
\cL = D + \P D^{-1} \Psi
\lab{aknslax}
\ee
The linear problem $\cL \psi_{BA} = \l \psi_{BA}$ can be decomposed as
\be
\pa_x \psi_{BA} + \P S (t, \l) =  \l \psi_{BA} \;\; ; \;\;
\pa_x  S (t, \l) = \Psi  \psi_{BA} (t, \l)
\lab{two-BA}
\ee
Similarly, we can introduce the conjugated linear problem:
$\cL^{*} \psi_{BA}^{*} =(- D - \Psi D^{-1} \Phi ) \psi_{BA}^{*}=
\l \psi_{BA}^{*}$ 
which can be rewritten as
\be
\pa_x S^{*} (t,\l) = \Phi \psi_{BA}^{*}
\;\; ; \; \; -\pa_x \psi_{BA}^{*}(t,\l) - \Psi S^{*} (t,\l) =  \l \psi_{BA}^{*}
\lab{conjl}
\ee
Recall from \ct{sep}, that in the Sato formalism the squared eigenfunction
potentials $S (t,\l), S^\ast (t,\l)$ are given by:
\br
S (t,\l) &=& {1 \o \l} \Psi (t-[\l^{-1}]) { \t (t-[\l^{-1}]) \o \t(t)}
 e^{ \xi (t,\l)} \lab{sep} \\
S^\ast (t,\l) &=& -{1 \o \l} \Phi (t+[\l^{-1}]) { \t (t+[\l^{-1}]) \o \t(t)}
e^{-\xi (t,\l)}\lab{sep-star} 
\er
where $ \xi (t,\l) = \sum \l^{j} t_j$.

We compare the above pseudo-differential setup to the algebraic dressing
formalism.
Consider, the  matrix $  L_0 \equiv D +E = D +\l \sigma_3 /2$
obtained by ``un-dressing'' the matrix Lax operator $L$ 
from eq.\rf{lopsl2}. Let
\br
\Psi^{+}_{vac} &= &e^{-\sum_i E^{(i)} t_i} \twocol{1}{0} =\twocol{1}{0}
e^{-\xi (t,\l)/2} \lab{psipvac}\\
\Psi^{-}_{vac} &= &e^{-\sum_i E^{(i)} t_i} \twocol{0}{1} =\twocol{0}{1}
e^{\xi (t,\l)/2} \lab{psimvac}
\er
be two solutions of equation $ L_0 \Psi_0=0 $.
Since, $ L_0 = S^{-1} U^{-1} L U S$ it follows that
$\Psi^{\pm} = US \Psi^{\pm}_{vac}= \Theta \Psi^{\pm}_{vac}$ 
satisfy $L\Psi^{\pm}=0$.

Let us write $\Theta$ as a $2 \times 2$ matrix:
\be
\Theta = \twomat{\th_{11}}{\th_{12}}{\th_{21}}{\th_{22}}
\lab{theta-mat}
\ee
then 
\br
L \Psi^{-}&=& \twomat{D+\l/2}{q(t)}{r(t)}{D-\l/2} \twocol{\th_{12}}{\th_{22}}
e^{\xi (t,\l)/2} =0 \nonu\\
\; \;& \to& \; \; \twomat{D}{q(t)}{r(t)}{D-\l} \twocol{\th_{12}}{\th_{22}}
e^{\xi (t,\l)} =0 
\lab{lpsiminus}
\er
The last equation can be cast into the form of 
\br
\th_{12} e^{\xi (t,\l)} &=& - \pa^{-1} \( q \th_{22} e^{\xi (t,\l)} \)
\lab{th12} \\
\l \th_{22} e^{\xi (t,\l)} &=& \( \pa - r \pa^{-1} q\) \th_{22} 
e^{\xi (t,\l)} 
\lab{th22}
\er
Comparing with \rf{two-BA} while making an identification 
\be
r = \Phi \qquad ; \qquad q = -\Psi
\lab{rpqp}
\ee
we find
\br
\th_{22} e^{\xi (t,\l)} &=& \psi_{BA}(t,\l) =  { \t (t-[\l^{-1}]) \o
 \t (t)} e^{\xi (t,\l)}
\nonu \\
\th_{12} e^{\xi (t,\l)} \!\! &=& \!\!- S \( q, \psi_{BA} (t,\l) \)= 
- {  q (t-[\l^{-1}]) \t (t-[\l^{-1}]) \o \l \t(t)}
e^{\xi (t,\l)}
\nonu 
\er
Similarly, 
\br
L \Psi^{+}&=& \twomat{D+\l/2}{q(t)}{r(t)}{D-\l/2} \twocol{\th_{11}}{\th_{21}}
e^{-\xi (t,\l)/2} =0 \nonu \\
& \to & \twomat{D+\l}{q(t)}{r(t)}{D} \twocol{\th_{11}}{\th_{21}}
e^{-\xi (t,\l)} =0 
\lab{lpsiplus}
\er
The last equation can be cast into the form of 
\br
\th_{21} e^{-\xi (t,\l)} &=& - \pa^{-1} \( r \th_{11} e^{-\xi (t,\l)} \)
\lab{th21} \\
\l \th_{11} e^{-\xi (t,\l)} &=& \( \pa - r \pa^{-1} q\)^{\ast}  
\th_{11} e^{-\xi (t,\l)} 
\lab{th11}
\er
Comparing with \rf{conjl} and \rf{rpqp} we find:
\br
\th_{11} e^{-\xi (t,\l)} &=& \psi_{BA}^{*}(t,\l) =  { \t (t+[\l^{-1}]) \o
 \t (t)} e^{-\xi (t,\l)}
\nonu \\ 
\th_{21} e^{-\xi (t,\l)} \!\!&=&\!\! - S \( r, \psi_{BA}^{*} (t,\l) \)= 
{ r (t+[\l^{-1}]) \t (t+[\l^{-1}]) \o \l \t(t)}
e^{-\xi (t,\l)}
\nonu 
\er
In this way we obtain the explicit matrix form of 
the matrices $\Theta $ and $\Theta^{-1}$ in
terms of the $\tau$ function:
\br
\Theta \!\!&=&\!\! {1 \o \t(t)}\twomat{ \t (t_{+}(\l)) }{
- {1\o \l}q (t_{-}(\l))) {\t (t_{-}(\l))}}{
{1\o \l} r (t_{+}(\l)) {\t (t_{+}(\l))}}{
{\t (t_{-}(\l))} }  \lab{Theta-tau} \\
\Theta^{-1}\!\! &=&\!\! {1 \o \t(t)}\twomat{ \t (t_{+}(\l)) }{
 {1\o \l}q (t_{-}(\l)) {\t (t_{-}(\l))}}{
-{1\o \l} r (t_{+}(\l)) {\t (t_{+}(\l))}}{
{\t (t_{-}(\l))} } \lab{Thetainv-tau} 
\er
with $t_{\pm} (\l) \equiv  t\pm [\l^{-1}]= (t_1 \pm 1/\l , t_2 \pm 1/2
\l^2,{\ldots} )$.
These expressions agree with the result of \ct{imbens} obtained within 
Wilson's framework \ct{W1,wilson81}.
The condition $\det \Theta =1$ implies:
\be
1= { \t (t+[\l^{-1}])  \t (t-[\l^{-1}])  \o  \t^2 (t)}
\(1+{ q (t-[\l^{-1}])  r (t+[\l^{-1}]) \o \l^2} \)
\lab{detone}
\ee
or, equivalently 
$\psi_{BA}(t,\l) \psi^\ast_{BA} (t,\l) + S(t,\l) S^\ast (t,\l) = 1$.

Writing $U$  as $U = \exp ( u_{+}(t, \l) \sigma_{+}+ u_{-}(t, \l) \sigma_{-})
 \exp ( \sg (t, \l) \sigma_3)$
and comparing with eq.\rf{Theta-tau} we obtain :
\be
\cosh^2 \(\sqrt{u_{+}u_{-}}\) =
\frac{\t (t-[\l^{-1}])\,\t (t+[\l^{-1}])}{\t^2 (t)} 
\lab{tau-alg-rel}
\ee
and
\be
e^{2\sg (\l)} = { \t (t+[\l^{-1}]) \o \t (t-[\l^{-1}]) }
\;\; \to \; \; \sg = \sum_{i=1}^{\infty} \sg^{(-i)}= 
\h \ln { \t (t+[\l^{-1}]) \o \t (t-[\l^{-1}]) }
\lab{sga}
\ee
or in terms of Schur polynomials :
\be
\sg^{(-n)} =  - {1  \o 2\l^{n} } \( p_n (- [\pa])  - p_n ( [\pa])  \)
\ln \t (t) \;\; ; \;\; n \geq 1
\lab{sgn}
\ee

\subsubsection{Hamiltonian Densities and the Tau Function of the AKNS Model}

In case of AKNS model quantities $\cH_n,\cJ_n$ become
\be
\cH_n = - \Tr \(\l^{n+1} {\sigma_3 \o 2} \Theta {\s_3 \o 2} \Theta^{-1} \)
\quad, \quad 
\cJ_n = - \Tr \(\l^{n+1}  \Theta_{\l} {\s_3 \o 2} \Theta^{-1} \)
\lab{chcj-akns}
\ee
where we introduced the notation $f_{\l}= d f = \l d f /d \l$.

Expressions \rf{Theta-tau} and \rf{Thetainv-tau} allow to calculate
\br
&&\Theta {\s_3 \o 2} \Theta^{-1} = -{1 \o 2} \s_3 \lab{usl2us3}\\
&+& {1 \o \t^2(t)} 
\twomat{\t (t_{+}(\l))  \t (t_{-}(\l))}
{{1\o \l} q (t_{-}(\l)) \t (t_{-}(\l))\t (t_{+}(\l))}
{ {1\o \l} r (t_{+}(\l)) \t (t_{-}(\l)) \t (t_{+}(\l))}
{\t (t_{+}(\l))\t (t_{-}(\l))}
\nonu
\er
which results in
\be
\Tr \( {\sigma_3 \o 2} \Theta {\s_3 \o 2} \Theta^{-1} \)
=  { \t (t+[\l^{-1}])  \t (t-[\l^{-1}])  \o  \t^2 (t)} - \h
\lab{trstst}
\ee
On the other hand from definition \rf{chcj-akns} we have :
\be
\Tr \( {\sigma_3 \o 2} \Theta {\s_3 \o 2} \Theta^{-1} \)
= - \sum_{n=1}^{\infty} \cH_n \l^{-n-1} +\h
\lab{genfunct}
\ee
Comparing the last two equations we find that the following must hold
\be
{ \t (t+[\l^{-1}])  \t (t-[\l^{-1}])  \o  \t^2 (t)}
= \sum_{n=1}^{\infty} {d \o d t_n} \pa_x \log (\t) \l^{-n-1} +1
\lab{hirota}
\ee
This is equivalent to the Hirota equations:
\be
\(  \h D_1 D_n - p_{n+1} ( [D]) \) \t \cdot \t =0
\lab{hiro-b}
\ee
due to identities:
\br
{ \t (t_{+}(\l))  \t (t_{-}(\l))  \o  \t^2 (t)}
&=& {1 \o \tau^2 \(t_i \)}
\exp \( \sum_{k=1}^{\infty} { \pa \o k \l^k \pa {\eps_k} }\) \t \(t + \eps \)
\t \(t - \eps  \) \v_{\eps=0}      \nonu \\
&=& {1 \o \t^2 \(t \)} \sum_{k=0}^{\infty} { p_k \( [{D}] \) \t \cdot \t \o \l^k}
\lab{psipsi} \\
{1 \o 2 \t^2 \(t \)} D_1 D_{n-1}
\t \cdot \t &=& \pa_x \pa_{n-1} \ln \t
\lab{sntau}
\er
where we used Hirota's operators defined by
\be
D^m_j a \cdot b = {\pa^m \o \pa s^m_j} a (t_j + s_j) b (t_j - s_j) \v_{s_j=0}
\lab{hiro-ope}
\ee
One can show that:
\br
\Tr \(  \Theta_{\l} {\s_3 \o 2} \Theta^{-1} \)
&=&  { 1  \o  2 \t^2 (t)} \( \t_{\l} (t_{+}(\l))  \t (t_{-}(\l)) 
 - \t (t_{+}(\l))  \t_{\l} (t_{-}(\l))\right. \nonu\\
&-&\left.
 {1\o \l^2} (q (t_{-}(\l)) \t (t_{-}(\l)))_{\l}
r (t_{+}(\l)) \t (t_{+}(\l)) \right. \nonu\\
 &+& \left. {1\o \l^2} q (t_{-}(\l)) \t (t_{-}(\l))
(r (t_{+}(\l)) \t (t_{+}(\l)) )_{\l} \)
\lab{tlsth}
\er
Observe, now that 
\be
f_{\l} (t \pm [\l^{-1}]) = \mp \sum_{k=0}^{\infty} \l^{-k} {d \o d t_k} f
(t\pm [\l^{-1}]) 
\lab{fltpl}
\ee
and therefore $\Tr \(  \Theta_{\l} {\s_3 \o 2} \Theta^{-1} \)$
can be rewritten as:
\br
&&{ - 1  \o  2 \t^2 (t)} 
\sum_{k=0}^{\infty} \l^{-k} {d \o d t_k}
 { \t (t_{+}(\l))  \t (t_{-}(\l)) }
\(1+{ q (t_{-}(\l))  r (t_{+}(\l)) \o \l^2} \) \nonu \\
&=& { - 1  \o  2 \t^2 (t)} 
\sum_{k=0}^{\infty} \l^{-k} {d \o d t_k} \t^2 (t)
\lab{detonea}
\er
where use was made of condition \rf{detone}.
We therefore find
\be
\Tr \(  \Theta_{\l} {\s_3 \o 2} \Theta^{-1} \) = - 
\sum_{k=0}^{\infty} \l^{-k} {d \o d t_k} \log \t (t)
\lab{tlstha}
\ee
in agreement with eq. \rf{ustau}.

Recall, that $\pa_x \sg^{(-1)} = -  k^{(-1)} = -qr$ and
$\theta^{(-1)} = u^{(-1)} + \sg^{(-1)}\sigma_3 $.
Accordingly,
\be
A^{(-1)} = {d \o d t_1} \theta^{(-1)} = \pa_x \theta^{(-1)}
= \pa_x \twomat{0}{-q}{r}{0} - qr \sigma_3
\lab{am1}
\ee
which is equal to :
\be
\(\Theta E {\Theta}^{-1} \)_{-1}
= \sbr{u^{(-2)}}{E^{(0)}} + \h \sbr{u^{(-1)}}{\sbr{u^{(-1)}}{E^{(0)}}}
\lab{thenthm1}
\ee
in agreement with \rf{crucial-proof}.

Inserting $ E^{(0)} = \sigma_3 /2$ and 
\be
A^{(-k)} = {d \o d t_k}  \theta^{(-1)}
= {d \o d\,  t_k}\( \twomat{0}{-q}{r}{0} + \sg^{(-1)} \sigma_3\)
\lab{amkk}
\ee
into the relation \rf{recura} we obtain (for $n >1$):
$$ 
{d \, \sg^{(-1)} \o d \, t_n}  = r {d \, q  \o d \, t_{n-1}}  -
q {d \, r \o d \, t_{n-1}}    + \sum_{k=1}^{n-2} \(
{d \, q \o d \, t_k}  {d \, r \o d \, t_{n-k-1}}  -
{d \, \sg^{(-1)} \o d \, t_k}    {d \, \sg^{(-1)}\o d \, t_{n-k-1}}  \)
$$ 
in agreement with reference \ct{slavnov}.

Recall, that $\cH_n= - d  \sg^{(-1)} / d t_n $.
Accordingly, the above equation becomes a recurrence relation for 
the Hamiltonian densities of the AKNS model :
$$ 
 \cH_n  = -r {d \, q  \o d \, t_{n-1}} + q {d \, r \o d \, t_{n-1}}   -
   \sum_{k=1}^{n-2} {d \, q \o d \, t_k}  {d \, r \o d \, t_{n-k-1}}
+2 rq {d\,   \sg^{(-1)} \o d \, t_{n-2}} +  \sum_{k=2}^{n-3}\cH_k \cH_{n-k-1}
$$ 
\subsection{Non-abelian Symmetries of the Integrable Models, sl(3) Example}
One of advantages of the dressing approach is that it provides a
convenient framework to classify and describe the symmetries of
integrable models.
In particular, the non-abelian symmetries emerge naturally in this framework 
for models with the non-abelian kernel $\cK$ of ${\rm ad}\, E$.
To illustrate the non-abelian symmetry structure of such models we consider 
here the linear spectral problem based on $sl (3)$ Lie algebra with 
the homogeneous gradation $\qs \equiv d$.
Here, the semi-simple 
and \underbar{non-regular} grade-one element $E$ is given by:
\be
E =  H^{(1)}_{\mu_2} = {\l \o 3} \threemat{1}{0}{0}{0}{1}{0}{0}{0}{-2} 
\lab{e-sl3}
\ee
Thus the kernel $\cK$ is the non-abelian sub-algebra
$\{ {\widehat {sl}} (2) \oplus {\hat U}(1) \}$ of 
$\cgh = {\widehat {sl}} (3)$ spanned by:
\be
\cK = \left\{ E^{(n)}\equiv \l^n H_{\mu_2},\, 
\l^n  H_{\mu_1},\, \l^n E_{\pm \a_1} \right\} 
\lab{ker-sl3}
\ee
where
\be
H_{\mu_1} = {1\o 3} \threemat{2}{0}{0}{0}{-1}{0}{0}{0}{-1}
\lab{hl1}
\ee
and in the Weyl basis
$E_{\a_1\,ij}= \d_{i1} \d_{j2}$ and $E_{- \a_1\,ij}= \d_{i2} \d_{j1}$.
The center $\cC (\cK)= \{ E^{(n)} \}={\hat U}(1) $ is spanned by
one element $H_{\mu_2}$ only.
The image is given by
$\cM = \left\{ E^{(n)}_{\pm \a_2},\, E^{(n)}_{\pm (\a_1 + \a_2)} \right\} $.
Accordingly, the Lax operator is:
\be
L = D\cdot I + E + A = D\cdot I + {\l \o 3}\threemat{1}{0}{0}{0}{1}{0}{0}{0}{-2} +
\threemat{0}{0}{q_1}{0}{0}{q_2}{r_1}{r_2}{0}
\lab{lax-sl3}
\ee
with the matrix $A \in \cM_0$.

The dressing procedure
\be
U^{-1} L U = D\cdot I +  {\l \o 3} \threemat{1}{0}{0}{0}{1}{0}{0}{0}{-2}
+ k_1 \l^{-1}+ O(\l^{-2})
\lab{ulusl3}
\ee
holds to the lowest order with 
\be
k_1 = \threemat{q_1r_1}{q_1r_2}{0}{q_2r_1}{q_2r_2}{0}{0}{0}{-q_1r_1-q_2r_2}
\lab{k1-sl3}
\ee
and 
\be
u^{(-1)} = \threemat{0}{0}{-q_1}{0}{0}{-q_2}{r_1}{r_2}{0} 
\lab{u1-sl3}
\ee
in 
$U= \exp ( u^{(-1)}\l^{-1} + O(\l^{-2}) )$.
We now apply these results to calculate
the symmetry transformations :
\be
\d^{(1)}_{\pm \a_1} A \equiv  \lb L \, , \, (\Theta \l^1 E_{\pm \a_1} \Theta^{-1})_{+} \rb 
= \lb \pa_x +  A\, , \, (\Theta \l^1 E_{\pm \a_1} \Theta^{-1})_{0} \rb 
\lab{symm-asl3}
\ee
where
\be
(\Theta \l^1 E_{\pm \a_1} \Theta^{-1})_{0} = 
\lb s^{(-1)}  \, , \, E_{\pm \a_1} \rb + 
\lb u^{(-1)}  \, , \, E_{\pm \a_1} \rb 
\lab{tetinv0}
\ee
and $ s^{(-1)} = - \pa^{-1} (k_1)$.
The transformations \rf{symm-asl3} are in components given by :
\br
\d^{(1)}_{\a_1} (q_1)\!\!\! &=&\!\!\! q_2^{\pr} -q_1 \pa^{-1} (q_2 r_1) -
q_2  \pa^{-1} (q_2 r_2 -q_1 r_1) \; ; \;
\d^{(1)}_{\a_1} (q_2) =  q_2 \pa^{-1} (q_2 r_1) \nonu \\
\d^{(1)}_{\a_1} (r_1)\! \!\!&=&\!\!\! r_1 \pa^{-1} (q_2 r_1) 
 \; ; \;
\d^{(1)}_{\a_1} (r_2) = r_1^{\pr} - r_2 \pa^{-1} (q_2 r_1) +
r_1  \pa^{-1} (q_2 r_2 -q_1 r_1) \nonu 
\er
and 
\br
\d^{(1)}_{-\a_1} (q_1)\!\!\! &=&\!\!\!  q_1 \pa^{-1} (q_1 r_2) \; ; \;
\d^{(1)}_{-\a_1} (q_2) = q_1^{\pr} -q_2 \pa^{-1} (q_1 r_2) +
q_1  \pa^{-1} (q_2 r_2 -q_1 r_1) \nonu \\
\d^{(1)}_{-\a_1} (r_1) \!\!\!&=&\!\!\! r_2^{\pr} - r_1 \pa^{-1} (q_1 r_2) -
r_2  \pa^{-1} (q_2 r_2 -q_1 r_1)
 \; ; \;
\d^{(1)}_{-\a_1} (r_2) = r_2 \pa^{-1} (q_1 r_2)  \nonu 
\er
These results can be reproduced compactly by a much simpler formula 
in the framework based on the pseudo-differential Lax operator.
To demonstrate this we note that the
matrix spectral problem $L \Psi =0 $ with $L$ from eq.\rf{lax-sl3}
can be reformulated in an equivalent form as the scalar spectral problem :
\be
\cL \psi_{BA} = \l \psi_{BA}\;\; ; \;\; \cL = D- r_1 D^{-1} q_1- r_2 D^{-1} q_2
\lab{BA-sl3}
\ee
Define :
\be
\cM_X  = \sum_{i,j=1}^2 X_{ij} r_i D^{-1} q_j
\lab{cmx-sl3}
\ee
for $X= E_{\pm \a_1}$, i.e. define $\cM_{E_{ \a_1}}  =  r_1 D^{-1} q_2$
and $\cM_{E_{- \a_1}}  =  r_2 D^{-1} q_1$.

We are now in position to reformulate transformations  \rf{symm-asl3} 
in one simple expression:
\be
\d^{(1)}_{\pm \a_1} \cL = - \sum_{i=1}^2 \d^{(1)}_{\pm \a_1} (r_i) D^{-1} q_i
- \sum_{i=1}^2 r_i D^{-1} \d^{(1)}_{\pm \a_1}  (q_i) \equiv
\lb \cM_{E_{\pm \a_1}} \, , \, \cL \rb
\lab{mxcl-sl3}
\ee
In calculating the left hand side of \rf{mxcl-sl3} we made use of identity:
\be
f_1 D^{-1} g_1 f_2  D^{-1} g_2 = f_1 \pa^{-1} (g_1 f_2) D^{-1} g_2-
f_1 D^{-1} g_2 \pa^{-1} (g_1 f_2)
\lab{iden-fgd}
\ee
By letting $X$ in eq. \rf{cmx-sl3} to be $\sigma_3$ and introducing higher
grade counterparts $\cL^n (r_i) , (\cL^{*})^n (q_i)$ of $ r_i, q_i$ 
we can extend the above results to obtain the graded Borel loop algebra of 
${sl} (2)$ within the pseudo-differential formalism.
See reference \ct{AGNP00} for details of this construction.

\sect{Additional Virasoro Symmetries}

\subsection{Virasoro Symmetry, the General Case}
We consider first the general case of the constrained KP models described by the
Lax operator $L=D_x +E+A$ within the ${\hat sl} (K+M+1)$ algebra
decomposed according to the grading operator $\qs$ from Section [\ref{background}].
The semisimple element $E$ of unit grade is given by \rf{a2}
while the potential $A$ is parametrized according to equation
\rf{a20}.

Define the modified ``bare'' Virasoro operators as
\be
X_{m(K+1)} = (K+1) l_m - \sum_{j=M+1}^{M+K} {\mu}_j \cdot H^{(m)} 
\lab{mod-vir}
\ee
where $\mu_a$ are fundamental weights of $sl (M+K+1)$ (as in Section
[\ref{background}]).
The operators $l_m =- \l^m d=-\l^{m+1} d/d \l$ satisfy the centerless Virasoro 
algebra \rf{vir-xmxn} :
\be
\sbr{l_m}{l_n} = (m-n) l_{m+n}
\lab{vir-xmxn}
\ee
For $b_N$ from $\cC (\cK)$ defined in \rf{a13} and $X_{N}$ from eq. \rf{mod-vir}
we find :
\be
\sbr{X_{N^{\pr}}}{b_N} = - N b_{N+N^{\pr}}
\lab{mira}
\ee
for $N^{\pr} = n (K+1)$. 
These relations imply that the modified Virasoro generators
${\ti X}_{N^{\pr}}$ defined as :
\be
{\ti X}_{m(K+1)} \equiv X_{m(K+1)} - \sum_I t_I b_{I+m(K+1)}
\lab{def-xtivir}
\ee
satisfy the centerless Virasoro algebra \rf{vir-xmxn}
with indices which are multiples of $K+1$  
\be
\sbr{{\ti X}_{m(K+1)}}{{\ti X}_{n(K+1)}} = (m-n)(K+1) {\ti X}_{(m+n)(K+1)}
\lab{vir-xmxn(K+1)}
\ee

Following equation \rf{symm-a} we define now the symmetry transformations
generated by the modified Virasoro generators ${\ti X}_m$ as :
\be
\d_m^V A = \lb D_x +E +A \, , \, ({\ti X}^{\Th}_m)_{+} \rb 
\lab{symm-avir}
\ee
This generates the Borel-Virasoro algebra which is also a symmetry of the model 
due to the fact that it commutes with the isospectral flows :
\be
\( \d_m^V {d \over d t_n}- {d \over d t_n}  \d_m^V \) {\Th} = 0
\quad , \quad m,n \geq 0
\lab{stani-tn}
\ee
The presence of the additional terms containing the time
parameters $t_I$ in definition \rf{def-xtivir} was crucial for commutativity
with isospectral times established in \rf{stani-tn}.

\subsection{The Homogeneous Gradation}
We now turn our attention to the additional Virasoro symmetry in 
case of homogeneous gradation.
Consider first the ``bare'' Virasoro operators 
$ X_m = l_m= - \l^m d \, ,\, m \geq 0$. 
in \rf{sym-flows} which satisfy the Witt algebra \rf{vir-xmxn}.

In that case the relation \rf{xde} no longer holds.
Instead one finds
\be
\sbr{l_m}{D_x+E}=- E^{(m+1)} 
\lab{lde}
\ee
as a special case of $\sbr{l_m}{b_n} = - n b_{m+n}$.
Relation \rf{lde} can be rewritten as
\be
\sbr{l_m-x E^{(m+1)}}{D_x+E}= 0
\lab{lede}
\ee
Applying ${\rm Ad}_{\Th}$ on \rf{lede} one finds the resolvent equation:
\be
\sbr{\Th \(l_m-x E^{(m+1)} \) \Th^{-1}}{\Th (D_x+E) \Th^{-1}}=0
\lab{resoL}
\ee
since $l_m-x E^{(m+1)} = \exp \(-x E \) l_m \exp \(x E \)$.

Define, now
\be
L_m = l_m - \sum_{i=1}^{\infty} i t_i  E^{(m+i)} \;\; ; \;\; {\rm for}
\;\;\;m \geq 0
\lab{extvira}
\ee
We are lead to:

\begin{definition}
Define a transformation $\d_m^V$ generated by
$ L_m$ from eq.\rf{extvira} as follows
\be
\d_m^V {\Th} \equiv (L^{\Th}_m)_{-} {\Th} \quad , 
\quad m \geq 0 
\lab{vir-flows}
\ee
\label{definition:virflows}
\end{definition}
where we defined 
\be
L^{\Th}_m \equiv {\Th} L_m {\Th}^{-1} = \Th  \(l_m- \sum_{i=1}^{\infty} i t_i E^{(m+i)} 
\) {\Th}^{-1}
\lab{defresolvv}
\ee

The  Witt algebra of the ``bare'' generators $L_m$ and $L_m$ 
results via relation \rf{dmdndk} for the Borel subalgebra of the Virasoro
algebra
\be
\( \d_m^V \d_n^V - \d_n^V \d_m^V \) {\Th} = (m-n) 
\d_{m+n}^V {\Th} \quad , \quad m,n \geq 0
\lab{stani}
\ee
\subsection{Example: Virasoro Symmetry of AKNS (Sl (2)) Hierarchy}

We now find action of the Virasoro symmetry on the
Lax coefficients $r, q$ from eq. \rf{lopsl2} describing the AKNS hierarchy
and compare with similar expressions found in the
formalism based on the pseudo-differential operators \ct{ourvirasoro}.

Virasoro transformations of $q,r$ are determined from:
\be
\d^V_n A = \sbr{L}{(L^{\Th}_n)_{+}} = \sbr{D+A}{(L^{\Th}_n)_{0}}
= - \sbr{E}{(L^{\Th}_n)_{-1}}
\lab{start}
\ee
for $L_n$ from eq. \rf{extvira}.
For the case of $\cG= sl(2)$,
$L^{\Th}_m= {\Th} L_m {\Th}^{-1}$ can be expressed as:
\be
L^{\Th}_n = U \( l_n - \sum_{j \geq 1} j \sg^{(-j)} \s_3 \l^{n-j} -  \sum_{k \geq 1}
k t_k E^{(k+n)} \) U^{-1}
\lab{ltm}
\ee
with $\sg^{(-1)} =  - \pa^{-1} (qr), \sg^{(-2)} = \h  \pa_2 \ln \t
= \h \pa^{-1} (r q_x -r_x q), {\ldots} $.
Recall also, that $E^{(k)}=b_k=\l^k \s_3/2$.

We now proceed by calculating $\d^V_n A $ from \rf{start} 
for $n=0,1,2$.

${\bf n=0}$. We find from \rf{ltm} that 
\br
(L^{\Th}_0)_0 &=& -d - \sum_{k \geq 1} k t_k (b_{k}^U)_0 \lab{lt0}\\
(L^{\Th}_0)_{-1} &=& -u^{(-1)}\l^{-1} - \sg^{(-1)}\l^{-1}- \sum_{k \geq 1} k t_k (b_{k}^U)_{-1}
\lab{ltm1}
\er
Plugging these two expressions into, respectively, \rf{start}
we find :
\be
\d^V_0 A = - A - \sum_{k \geq 1} k t_k {d A \o d t_k}
\lab{dvza}
\ee
or 
\be
\d^V_0 r  = - r - \sum_{k \geq 1} k t_k {d r \o d t_k} \lab{dvzr} \;\; ; \;\; 
\d^V_0 q  = - q - \sum_{k \geq 1} k t_k {d q \o d t_k} \lab{dvzq} 
\ee
${\bf n=1}$. We find from \rf{ltm} that 
\br
(L^{\Th}_1)_0 &=& -(u^{(-1)} + \sg^{(-1)}\s_3 ) - 
\sum_{k \geq 1} k t_k (b_{k+1}^U)_0 \lab{ltone0}\\
(L^{\Th}_1)_{-1} &=& -2 \l^{-1} (u^{(-2)} + \sg^{(-2)}\s_3)-\l^{-1}
\sbr{u^{(-1)}}{\sg^{(-1)}\s_3}
\lab{ltonem1}\\
&-& \sum_{k \geq 1} k t_k (b_{k+1}^U)_{-1} \nonu
\er
which lead via \rf{start} to :
\br
\d^V_1 r & =& - 2r_x - 2 r (\ln \t)_x - \sum_{k \geq 1} k t_k {d r \o d
t_{k+1}} \lab{dv1r} \\
\d^V_1 q & =& 2 q_x +2 q (\ln \t)_x - \sum_{k \geq 1} k t_k 
{d q \o d t_{k+1}} \lab{dv1q} 
\er
${\bf n=2}$. This time we find from \rf{ltm} that 
\br
(L^{\Th}_2)_0 \!\!&=&\!\! -2 (u^{(-2)} +\sg^{(-2)}\s_3)- 
\sbr{u^{(-1)}}{\sg^{(-1)}\s_3}
- \sum_{k \geq 1} k t_k (b_{k+2}^U)_{0}
\nonu \\
(L^{\Th}_2)_{-1} \v_{\cM} \!\!&=&\!\! -3 \l^{-1} u^{(-3)} -\l^{-1}
\sbr{u^{(-2)}}{\sg^{(-1)}\s_3}
 \nonu\\
\!\!&-&\!\! 2 \l^{-1} \sbr{u^{(-1)}}{\sg^{(-2)}\s_3}
  - \sum_{k \geq 1} k t_k (b_{k+2}^U)_{-1} \v_{\cM}
\lab{lttwom1}
\er
Plugging expression from \rf{lttwom1} into \rf{start} we obtain:
\br
\d^V_2 r \!\!\!& =&\!\!\! - 3r_{xx} - 2 r_x (\ln \t)_x -2 r \pa_2 (\ln \t)
+4qr^2 - \sum_{k \geq 1} k t_k {d r \o d
t_{k+2}} \lab{dv2r} \\
\d^V_2 q \!\!\!& =&\!\!\! -3 q_{xx} -2 q_x (\ln \t)_x +2 q \pa_2 (\ln \t)
+4 q^2 r - \sum_{k \geq 1} k t_k 
{d q \o d t_{k+2}} \lab{dv2q} 
\er
The crucial observation is that
the transformation:
\be 
\d^V_n \; \to \; {\ti \d}^V_n \equiv \d^V_n + {(n+1) \o 2 } {d \o d t_n}
\lab{tid}
\ee
preserves the Virasoro algebra, meaning that ${\ti \d}^V_n $ satisfies 
the Virasoro algebra.
Taking into account that ${d / d t_n}$ is generated by $B_n$
one obtains the following expressions:
\br
{\ti \d}^V_0 r \!\!\!& =& \!\!\!- r/2 - \sum_{k \geq 1} k t_k {d r \o d t_k}
\lab{tidvzr}\\ 
{\ti \d}^V_1 r \!\!\!& =& \!\!\!- r_x - 2 r (\ln \t)_x - \sum_{k \geq 1} k t_k {d r \o d
t_{k+1}} \lab{tidv1r} \\
{\ti \d}^V_2 r \!\!\!&=&\!\!\! - {3 \o 2} r_{xx} - 2 r_x (\ln \t)_x -2 r \pa_2 (\ln \t)
+qr^2 - \sum_{k \geq 1} k t_k {d r \o d
t_{k+2}} \lab{tidv2r} 
\er
We will now attempt to rewrite the above relations in the Sato
pseudo-differeential Lax formalism.
For this purpose we need to introduce Orlov-Shulman operator $M$ in 
addition to the Lax operator $\cL= D- r D^{-1} q = D+ \P D^{-1} \Psi$.
$M$ is defined in such a way that
\be
M \psi_{BA} (t, \l) = \partder{}{\l} \psi_{BA} (t, \l)
\lab{monpsi}
\ee
for the Baker-Akhiezer wave function :
\be
\ln \psi_{BA} (t, \l) = \sum_{n=1}^{\infty} t_n \l^n + \sum_{n=1}^{\infty}
\l^{-n} p_n (- [\pa]) \ln \t 
\lab{lnpsi}
\ee
and therefore Orlov-Shulman operator $M$ can be written as
\be 
M = \sum_{n=1}^{\infty} n t_n \cL^{n-1} + \sum_{n=1}^{\infty}
 \( -n p_n (- [\pa]) \ln \t  \)\, \cL^{-n-1}
\lab{Mlax}
\ee
Note, that since $\cL \psi_{BA} (t, \l) = \l \psi_{BA} (t, \l)$ we have 
$ \lb \cL \, , \, M \rb =1$

Using representation of the Orlov-Shulman operator given above in equation \rf{Mlax}
and identity $\pa_n \pa_x \ln \tau = - {\rm Res} (\cL^n)$
for $n=1,2$ we can rewrite relations \rf{tidvzr}-\rf{tidv2r} as :
\br
{\ti \d}^V_0 \P & =& - \P/2 - \( M \cL\)_{+} (\P)
\lab{tidvzP}\\ 
{\ti \d}^V_1 \P & =& - \cL (\P) - \( M \cL^2\)_{+} (\P)
\lab{tidv1P} \\
{\ti \d}^V_2 \P &=& - {3 \o 2} \cL^2 (\P)- X^{(1)}_2 (\P)- (M\cL^3)_{+}(\P)  
 \lab{tidv2P} 
\er
where use was made of identification of $r,q$ with $\Phi, -\Psi$ (see f.i.
\rf{rpqp}) and where the pseudo-differential object $X^{(1)}_2$
\be
X^{(1)}_2 = - \h \cL (\P) D^{-1} \Psi + \h \P D^{-1} \cL^* ( \Psi)
\lab{xonetwo}
\ee
is a special case of
\be
X^{(1)}_k = \sum_{j=0}^{k-1} \lb j - \h (k-1)\rb \cL^{k-1-j} (\P) D^{-1}
(\cL^*)^j (\Psi)
\lab{xdef}
\ee
All these results in \rf{tidvzP}-\rf{tidv2P} agree perfectly well with 
reference \ct{ourvirasoro} (up to an overall minus sign).

Now we deal with action of Virasoro transformations
${\ti \d}^V_n \equiv \d^V_n + {(n+1) \o 2 } {d \o d t_n}$
from \rf{tid} applied on $q$.

Recalling that $q$ is an adjoint eigenfunction i.e. $ \pa q /d t_n = - B^*_n (q)$
we obtain the following expressions:
\br
{\ti \d}^V_0 q \!\!\!& =&\!\!\! - 3q/2 - \sum_{k \geq 1} k t_k {d q \o d t_k} 
\lab{bardvzq} \\
{\ti \d}^V_1 q \!\!\!& =&\!\!\!  3 q_x +2 q (\ln \t)_x - \sum_{k \geq 1} k t_k 
{d q \o d t_{k+1}} \lab{bardv1q} \\
{\ti \d}^V_2 q \!\!\!&=&\!\!\! - {9 \o 2} q_{xx} - 2 q_x (\ln \t)_x +2 q \pa_2 (\ln \t)
+7 q^2r - \sum_{k \geq 1} k t_k {d q \o d t_{k+2}} \lab{bardv2q} 
\er
Since
\be
\ln \psi_{BA}^* (t, \l) = -\sum_{n=1}^{\infty} t_n \l^n + \sum_{n=1}^{\infty}
\l^{-n} p_n ( [\pa]) \ln \t 
\lab{lnpsistar}
\ee
we find that $M^*$ such that:
\be
M^* \psi_{BA}^* (t, \l) = - {d \o d \l} \psi_{BA}^* (t, \l) 
\lab{cmstar}
\ee
is equal :
\be
M^* =\sum_{n=1}^{\infty} n t_n (\cL^{*})^{n-1}  + \sum_{n=1}^{\infty}
 \( n p_n ( [\pa]) \ln \t  \)\, (\cL^{*})^{-n-1}
\lab{Mlaxadj}
\ee
and satisfies $ \lb  M^* \, , \, \cL^*  \rb =1$.

With this definition and 
identification \rf{rpqp} relations \rf{bardvzq}-\rf{bardv2q}  take form
\br
{\ti \d}^V_0 \Psi & =& - \Psi/2 + \( M\cL\)_{+}^* (\Psi)
\lab{barvzPsi}\\ 
{\ti \d}^V_1 \Psi & =& - \cL^* (\Psi) + \( M \cL^{2}\)_{+}^*  (\Psi)
\lab{barv1Psi} \\
{\ti \d}^V_2 \Psi &=& - {3 \o 2} (\cL^{*})^{2} (\Psi)+ X^{(1)\, *}_2 (\Psi)
+ (M\cL^{3})_{+}^* (\Psi)  
 \lab{barv2Psi} 
\er
with 
\be
X^{(1)\, *}_2 =  \h \Psi D^{-1}\cL (\P)  - \h \cL^* ( \Psi)D^{-1}\P 
\lab{xonetwostar}
\ee

Relations \rf{barvzPsi}-\rf{barv2Psi}
again agree with 
the reference \ct{ourvirasoro} (up to an overall minus sign).

\sect{Fermionic Symmetry Flows from the Super Algebra}

Consider the super-algebra $A(p,s)$ composed of the bosonic sub-algebra
\be
SL(p+1)\otimes SL(s+1)\otimes U(1)
\lab{aps}
\ee
together with the fermionic generators 
$ E_{\pm (\a_i +\a_{i+1} +\cdots +\a_{p+1} +\a_{p+2} + \cdots +\a_j)}$, $ i=1,
\cdots, p+1$, $j= p+1, \cdots, p+s+ 1$.

The root system can be realized in terms of $p+s+1$ orthonormal vectors,
$e_i \cdot e_k = \d_{ik} $, $ f_j \cdot f_l = -\d_{jl}$, 
$i,k = 1, \cdots ,p+1$, $j,l = 1, \cdots ,s+1$.
Let $\a_1 = e_1 - e_2, \a_2 =  e_2-e_3, \cdots , \a_p =e_p -e_{p+1}$ and $
\a_{p+2}= f_{1}-f_{2}, \cdots, \a_{p+s+1}= f_{s}- f_{s+1}$ be the
bosonic simple roots of $SL(p+1)\otimes SL(s+1)$ and denote by $\a_{p+1} =
e_{p+1} - f_1$ the fermionic simple root.

We now define $E= h_{p+1}^{(1)}=\a_{p+1} \cdot H^{(1)}$ as the constant 
Lie algebra valued element with unit grade with respect to the homogeneous 
grading $Q=d$ which defines decomposition of the super Lie algebra
$A(p,s)$ on kernel and image of $\mbox{ad E}$.  The grade zero part of the
kernel is :
\br
{\kere}_0  &=& \{ SL(p) \otimes SL(s) \otimes
U(1)^3 \} \oplus \{ E^{(0)}_{\pm \a_{p+1}}, 
E^{(0)}_{\pm (\a_p +\a_{p+1} +\a_{p+2})},
\nonu \\
&\ldots& 
,E_{\pm (\a_i + \a_{i+1} +\cdots +\a_{p+1} + \cdots +\a_{p+j})} \}
\lab{ker-fer}
\er
where $ i=1, \ldots , p+1, \; j=1, \ldots, s+1$ and
$U(1)^3$ is generated by $h_{p+1}$ and
$ \mu_p \cdot H, \mu_{p+2} \cdot H$.
The presence of the fermionic
generators $$E_{\pm (\a_i + \cdots + \a_{p+1} +  \cdots + \a_{p+j})}$$ in
$\kere$ is a consequence of the indefinite metric as expressed by
$e_i \cdot e_k = \d_{ik}, f_k \cdot f_l = -\d_{kl}$.

The center of $\kere$ is generated by $h_{p+1}$ and is related with 
the bosonic flows.

In addition, we introduce fermionic elements $F_{\pm} \equiv
E_{\a_{p+1}}^{(0)} \pm E_{-\a_{p+1}}^{(1)}$ from eq.\rf{ker-fer}
whose squares reproduce the unit grade constant element according to
$ \h \{  F_{\pm}\, , \, F_{\pm}\} =  \pm h_{p+1}^{(1)} = \pm E$.
Moreover it holds that $\{  F_{\pm}\, , \, F_{\mp}\}=0$.
We note that according to gradation $Q^{\pr} = 2 d + \mu_{p+1} \cdot H$
the elements $F_{\pm}$ possess the unit grade.

We can generalize these definitions to fermionic elements of grade $n$ with
respect to grading defined by $Q^{\pr}$ as
$F_{\pm}^{[n]} \equiv
E_{\a_{p+1}}^{(n)} \pm E_{-\a_{p+1}}^{(n+1)}\, , \, n\geq 0$.
They satisfy the anti-commutation relations
$\h \{  F_{\pm}^{[n]}\, , \, F_{\pm}^{[m]}\} = \pm h_{p+1}^{(m+n+1)}  $
and $\{  F_{\pm}^{[n]}\, , \, F_{\mp}^{[m]} \}=0$.

We can now associate symmetry flows to the elements $F_{\pm}^{[n]} $
within our dressing framework.
This is illustrated in the following example.

\subsection{Example of Super-algebra ${\bf sl(2|1)}$}

The super-algebra $sl(2|1)$ is the (N=2) extended
supersymmetric version of $sl(2)$ and contains four bosonic generators
$E_{\a_1},E_{-\a_1},H_1,H_2$ which form the Lie
algebra $sl(2) \oplus U(1)$ and four fermionic generators
$E_{\a_2},E_{-\a_2},E_{\a_1+\a_2}$ and $E_{-\a_1-\a_2}$.
Here $\a_1=e_1-e_2$ and $\a_2=e_2-f_1$ denote simple bosonic and fermionic
roots. 
In this notation we have the following Cartan elements
\be
{\a_2} \cdot H = H_1+H_2 \quad ; \quad 
-({\a_1+\a_2}) \cdot H = H_1-H_2 \quad ; \quad 
-{\a_1} \cdot H = 2 H_1
\lab{cartans}
\ee
The three-dimensional matrix representation (fundamental 
representation) of the above operators in the 
Cartan-Weyl basis reads as
\begin{eqnarray*}
&& H_1 = \left( \begin{array}{ccc} 
\h & 0 & 0 \cr 0 & -\h & 0 \cr 0 & 0 & 0
\end{array} \right) \quad
H_2 = \left( \begin{array}{ccc} 
\h & 0 & 0 \cr 0 & \h & 0 \cr 0 & 0 & 1
\end{array} \right) \quad
E_{-\a_1} = \left( \begin{array}{ccc} 
0 & 1 & 0 \cr 0 & 0 & 0 \cr 0 & 0 & 0
\end{array} \right) \ \\
&& E_{\a_1} = \left( \begin{array}{ccc} 
0 & 0 & 0 \cr 1 & 0 & 0 \cr 0 & 0 & 0
\end{array} \right) \quad
E_{\a_2} = \left( \begin{array}{ccc} 
0 & 0 & 1 \cr 0 & 0 & 0 \cr 0 & 0 & 0
\end{array} \right) \quad
E_{-\a_1-\a_2} = \left( \begin{array}{ccc} 
0 & 0 & 0 \cr 0 & 0 & 0 \cr 0 & 1 & 0
\end{array} \right) \\
&&E_{-\a_2} = \left( \begin{array}{ccc} 
0 & 0 & 0 \cr 0 & 0 & 0 \cr 1 & 0 & 0
\end{array} \right) \quad
E_{\a_1+\a_2} = \left( \begin{array}{ccc} 
0 & 0 & 0 \cr 0 & 0 & 1 \cr 0 & 0 & 0
\end{array} \right)
\end{eqnarray*}
The model is considered with the homogeneous gradation.
Furthermore, a semisimple grade-one element is taken to be
$E= \l \a_2 \cdot H= \l (H_1+H_2)= \l {\rm diag} (1,0,1)$.
Accordingly, the corresponding kernel $\cK$
\be
\cK= \kere = \{ \l^n H_1, \l^n H_2, \l^n E_{\a_2}, \l^n  E_{-\a_2} \}
\lab{kersl21}
\ee
contains fermionic roots. We consider an element $F= E_{\a_2}^{(0)}+  E_{-\a_2}^{(1)}$
such that $F^2=E$. The role of $F$ was recognized already in \ct{AD98}
in construction of the fermionic Lax operator for s-AKNS hierarchy.
Note, that according to gradation $Q^{\pr} = 2 d + H_1-H_2 $
the element $F$ possesses a unit grade.

The higher grade generalizations of $F$ defined as
$F_{\pm}^{[n]} \equiv
E_{\a_{2}}^{(n)} \pm E_{-\a_{2}}^{(n+1)}$ for $ n\geq 0$
are in $\cK$ and have grade $2n+1$ with respect to grading defined by $Q^{\pr}$. 
They satisfy the commutation relations
\be
 \{  F_{\pm}^{[n]}\, , \, F_{\pm}^{[m]}\} = \pm 2 (H_1+H_2)^{(m+n+1)}  
= \pm 2 E^{(m+n+1)}  
\lab{n1a}
\ee
and 
\be
\{  F_{\pm}^{[n]}\, , \, F_{\mp}^{[m]} \}=0
\lab{ffpm}
\ee
In addition, we will also list commutation relations of $F_{\pm}^{[n]}$
with another (apart from $E$) Cartan operator $H_1-H_2$ from $\cK$:
\be
\{   (H_1-H_2)^{(n)} \, , \, F_{\pm}^{[m]}\} = - F_{\mp}^{[n+m]}
\lab{nmax}
\ee
We encounter  here an example of the model which contains
fermionic elements in $\kere$.
Accordingly, the above algebraic structure will give rise to the graded algebra of 
flows as follows.
Let the bosonic Lax $L=D +E + A$ be given with 
the potential $A$, as usually, determined by the condition that all its
components are in the zero-grade subspace of $\cM=\ime$ \ct{ADR97} :
\be
A= b_1  E_{-\a_1}+ b_2  E_{\a_1} + f_1  E_{\a_1+\a_2}+
f_2  E_{-\a_1-\a_2}
 =  \threemat{0}{b_1}{0}{b_2}{0}{f_1}{0}{f_2}{0}
\lab{asl21}
\ee
Next, we associate the symmetry flows $\partial / \partial \tau_n^{\pm}$ 
to the the odd elements  $F_{\pm}^{[n]}$ according to the definition:
\be
{\partial \over \partial \tau_n^{\pm}} \Theta \,= \, \d^{(n)}_{F_{\pm}} \Theta \,=\, 
(\Theta F_{\pm}^{[n]} \Theta^{-1})_{-} \Theta
\lab{fern}
\ee
which furthermore are assumed to anti-commute with fermionic roots and so
\be
{\partial \over \partial \tau_n^{\pm}} F_{\pm}^{[m]} = - F_{\pm}^{[m]} {\partial \over \partial \tau_n^{\pm}}
\lab{antfm}
\ee
Next, we associate the symmetry flows $\partial / \partial u_n$ to 
Cartan operator $(H_1-H_2)^{(n)}$ via:
\be
{\partial \over \partial u_n} \Theta \,= \, (\Theta (H_1-H_2)^{(n)} \Theta^{-1})_{-} \Theta
\lab{h1ern}
\ee
{}From equations   \rf{dmdndk} and \rf{com-iso} and eq.\rf{n1a}
we find that the
fermionic flows commute with isospectral flows and close into the
isospectral flows generated by $E^{(n)}$ as follows
\be
\( {\pa \over \pa \tau_m^{\pm}} {\pa \over \pa \tau_n^{\pm}} +  
 {\pa \over \pa \tau_n^{\pm}} {\pa \over \pa \tau_m^{\pm}} \) \Theta
 = \pm 2 {\pa \over  \pa t_{m+n+1} }\Theta \quad , \quad m,n \geq 0
\lab{ferm-tn}
\ee
and satisfy in addition
\be
\( {\pa \over \pa u_m} {\pa \over \pa \tau_n^{\pm}} -
 {\pa \over \pa \tau_n^{\pm}} {\pa \over \pa u_m} \) \Theta
= - {\pa \over \pa \tau_{m+n}^{\mp}} \Theta
   \quad , \quad m,n \geq 0
\lab{uerm-tn}
\ee
Notice that algebra of the flows $\partial / \partial \tau_n^{-}$ is
isomorphic to the Manin-Radul algebra of flows.
The extended algebra of flows $\partial / \partial \tau_n^{\pm}$, ${\pa \over \pa u_m}$
together with isospectral flows has been encountered in the study
of the maximal SKP hierarchy (see e.g. \ct{tak}, \ct{sorin}).

We have shown above that the model possesses additional fermionic symmetry
flows. Let us now find their explicit form.
Via the dressing technique we arrive at
\be
\d^{(1)}_{F} A \equiv  \lb L \, , \, (\Theta F \Theta^{-1})_{+} \rb 
= \lb \pa_x +  A\, , \, (\Theta F \Theta^{-1})_{0} \rb 
\lab{symm-F}
\ee
Working out the lowest terms $u^{(-1)}, \sg^{(-1)}$ in the grading expansion
of $\Theta$ and plugging them into expression for $(\Theta F \Theta^{-1})_{0}$
in \rf{symm-F}
we obtain for $b_1$ and $f_1$ components of $A$ in 
\rf{asl21} the following transformations :
\be
\d^{(1)}_{F} b_1 = -f_2 + b_1 \int b_1 f_1 \quad ; \quad
\d^{(1)}_{F} f_1 = b_2 + f_1 \int b_1 f_1 
\lab{dfbf}
\ee
To understand these flows and their connection to supersymmetry we recall from
\ct{ADR97} that the
matrix spectral problem $L \Psi =0 $ with $L=D+E+A$
can be reformulated in the equivalent form of the scalar spectral problem :
$\cL \psi_{BA} = \l \psi_{BA}$ 
with the pseudo-differential Lax operator:
\be 
\cL = D + \Phi (t, \th ) 
{D_{\th}}^{-1} \Psi (t, \th ) 
\lab{eigenl}
\ee
where the superfields $\Phi (t, \th )$ and
$\Psi (t, \th)$ are, respectively,
eigenfunctions and  adjoint  eigenfunctions
of $\cL$ and $D_{\th} $ is a covariant derivative of the form:
$ D_{\th} = \partder{}{\th} + \th \pa$, which
satisfies $ {D_{\th}}^2 = \pa$.

The paper \ct{ADR97} established the following connection between components
of $A$ and the superfields $\Phi (t, \th )$ and
$\Psi (t, \th)$ :
\be
b_1=\Phi (t, \th ) \; , \; f_1 = \Psi (t, \th)\; , \; 
b_2= -D_{\th} \Psi + (\int \Phi \Psi) \Psi \; , \; 
f_2 = D_{\th} \Phi + (\int \Phi \Psi) \Phi  
\lab{sup-dict}
\ee
Inserting these values into the transformation law 
\rf{dfbf} we find that
\be
\d^{(1)}_{F} \Phi (t, \th ) = - D_{\th}  \Phi (t, \th ) \; \; , \;\;  
\d^{(1)}_{F}\Psi (t, \th) = - D_{\th}  \Psi (t, \th )
\lab{dfPPsi}
\ee
Hence the first flows associated to $F$ amount to application of the covariant
derivative.
In order to find the higher flows $\d^{(2n+1)}_{F}$ generated by
$ F_{+}^{[n]}$ we employ the recursion techniques from \ct{ADR97}
generalized to odd/half-integer flows $\pa / \pa t_{2n+1}
\equiv \d^{(2n+1)}_{F}$ entering the zero curvature equation:
\be
{\pa \o \pa t_{2n+1}} A - \pa B_{2n+1} + \l \lb E \, , \, B_{2n+1} \rb
+\lb A \, , \, B_{2n+1} \rb = 0
\lab{zsa}
\ee
with
\be
B_{2n+1} = F_{+}^{[n]}+B_n + {\ldots} +B_0
\lab{b2n1-exp}
\ee
where terms $B_k$ have grade equal to $k$.
After plugging expansion \rf{b2n1-exp} into relation \rf{zsa} and decomposing  
it according to the grade we find
\be
\d^{(2n+1)}_{F} (A_E) = (-\cR)^n \( \d^{(1)}_{F} (A_E) \)
\lab{dfnae}
\ee
where $A_E \equiv ad_{E} (A) $ and the recursion matrix is given by :
\be
\cR \equiv  ad_{E} \( \pa - ad_{\cA}\, \pa^{-1} ad_{\cA} \) 
\lab{recoper}
\ee

\sect{Background on Graded Affine Lie Algebras}
\label{background}
In this section we provide the basic ingredients 
about the graded affine Lie algebras needed in 
construction of integrable hierarchies of the constrained KP type,
for more details see \ct{AFGZ-jmp} and references therein.

Let $\cgh$ be an affine Lie algebra, and  $\lie$ be the
 finite dimensional simple Lie algebra associated to it.
The integral gradation of $\cgh$ defines the following decomposition :
\be
\cgh = \bigoplus_{n\in \IZ} \cgh_n(\bfs )\, , \qquad
\lb \cgh_m(\bfs ) \, , \, \cgh_n(\bfs )\rb \subset \cgh_{m+n}(\bfs )
\ee
where $\cgh_n(\bfs )$ is a grade-$n$ subspace:
\be
\lb \qs \, , \, \cgh_n(\bfs )\rb = n\, \cgh_n(\bfs )  
\lab{grade-eigenspace}
\ee
with respect to the grading operator :
\be
\qs \equiv \sum_{a=1}^{r} s_a\, {2 \mu_a \cdot H^0\over \a_a^2}
+ N_{\bfs} d
\lab{grading}
\ee
The following ingredients entered the definition \rf{grading}.
The vector $\bfs = \( s_0,s_1, \ldots, s_r\)$ \ct{kac}, has 
components $s_i$ being non
negative  relatively  prime integers, and $r\equiv {\rm rank}\, \lie$.
Moreover, $H^0_a$, $a=1,2,\ldots, r$, are the Cartan sub-algebra generators
of $\lie$, $\mu_a$ its fundamental weights satisfying
${2 \mu_a \cdot \a_b \over \a_b^2}= 2 \d_{ab}$, with $\a_a$ being the
simple roots of $\lie$.
$d= \l d /d \l$ is the usual derivation of $\cgh$, responsible for
the homogeneous gradation of $\cgh$, corresponding to
$\bfs_{\rm hom}= (1,0,0,\ldots ,0)$. In addition, we have,
$N_{\bfs} \equiv \sum_{i=0}^{r} s_i m_i^{\psi}$,
$\psi = \sum_{a=1}^{r}  m_a^{\psi} \a_a$, $m_0^{\psi} = 1$,
where $\psi$ is the highest positive root of $\lie$.

\subsection{The case of $\cgh={\bf {\widehat {sl}}(M+K+1)}$}
\label{slmk}

We now apply the above formalism
to the example of the affine Lie algebra
$\cgh = {\widehat {sl}} \( M+K+1 \)$, ($A_{M+K}^{(1)}$)
furnished with gradation $\bfs$ and corresponding grading operator $\qs$ :
\be
\bfs = ( 1, \underbrace{0, \ldots ,0}_{M}, \underbrace{1, \ldots,1}_{K}\, )
\;\; ;\;\; \qs = \sum_{j=M+1}^{M+K} \mu_j \cdot H^{(0)} + (K+1) d
\lab{a1}
\ee
We will denote the simple roots of ${\widehat{sl}} \( M+K+1 \)$ by $\a_j$,
$j=0,1,\ldots, M+K$, with $\a_0 \equiv -\psi$ for $\psi$ being the highest
positive root of $\lie = sl \( M+K+1 \)$.
All roots are such that $\a_j^2=2$.

The semisimple, grade-one (w.r.t. to gradation $\bfs$) element $E$ is taken
to be :
\be
E = \sum_{j=M+1}^{M+K}  E^{(0)}_{\a_{j}}
+  E^{(1)}_{-(\a_{M+1}+ \cdots+\a_{M+K})}
\lab{a2}
\ee
it's centralizer is :
\be
\cK= 
\kere = \{ {\hat K}_0 \equiv {\widehat{sl}} (M) \oplus {\hat U} (1) \; , \;
{\hat {\cH}}_K \}
\lab{a8}
\ee
where ${\widehat{sl}} (M)$ is the affine Lie sub-algebra of
$\cgh = {\widehat{sl}} (M+K+1)$ with simple roots  $\a_j$, $j=1, 2, \ldots,
M-1$ and $\a_0=-(\a_1 + \a_2 + \ldots + \a_{M-1})$. The algebra
${\hat U} (1)$ is generated by $\mu_M \cdot H^{(k)}$, $k\in \IZ$.
In addition, ${\hat{\cH}}_K$ is the sub-algebra of
${\widehat{sl}} (K+1) \in {\widehat{sl}} (M+K+1)  $ 
and spanned by generators :
\br
E_{l+(K+1)n} \eq E^{(n)}_{\a_{M+1}+\a_{M+2}+\ldots +\a_{M+l}} +
E^{(n)}_{\a_{M+2}+\a_{M+3}+\ldots + \a_{M+l+1}} + \ldots
\nonu \\
&+&
E^{(n)}_{\a_{M+K-l+1} + \a_{M+K-l+2} +\ldots +\a_{M+K-1}+\a_{M+K}} \nonu \\
&+&
E^{(n+1)}_{-(\a_{M+1}+\a_{M+2}+\ldots +\a_{M+K-l+1})} +
E^{(n+1)}_{-(\a_{M+2}+\a_{M+3}+\ldots +\a_{M+K-l})}  
\nonu \\
&+& \ldots +
E^{(n+1)}_{-(\a_{M+l} + \a_{M+3}+\ldots +\a_{M+K})}
\er
with $l=1, 2, 3, \ldots,K$. Note, that $E_1 = E$.
These generators satisfy
\be
\sbr{\qs}{E_{l+(K+1)n}} = (l+(K+1)n)\, E_{l+(K+1)n}
\ee
Also, we have
\be
\cC (\cK) = {\rm center }\, \kere = \{  {\hat U} (1) \; , \;{\hat{\cH}}_K \}
\lab{a13}
\ee
where ${\hat U} (1) $ is as in eq.\rf{a8}.
Notice that
$\sbr{\qs}{\mu_M \cdot H^{(k)}} = k (K+1) \mu_M \cdot H^{(k)}$.
The center of $\kere$ has one and only one generator associated to a given
grade according to the scheme:
\br
b_{N} &=&  E_{N=l+(K+1)n} 
\qquad l =1,2, \ldots, K \lab{a16a} \\
b_{k(K+1)} &=& \mu_M \cdot H^{(k)}\, , \, \qquad k \in \IZ
\lab{a16b}
\er
According to \rf{sym-flows}, each of the generators from the center of 
\kere in \rf{a16a}-\rf{a16b}
will give rise to the corresponding isospectral flows with
times $t_{b_{N}}, t_{b_{k(K+1)}}$.
In particular the element $E_1 = E$
will generate the flow corresponding to $\pa / \pa t_1 = \pa / \pa x$.

The generators of the complement $\cM$ of $\cK$ within the grade zero
sub-algebra $\cgh_0$ are :
\be
\cM_0 = \{ P_{\pm i} = E_{\pm (\a_{i} + \a_{i+1} + \ldots +\a_{M})}^{(0)}\,
,\,\a_a \cdot H^{(0)} \}
\lab{pis}
\ee
for $i=1, 2, \ldots, M$ and $a=M+1, \ldots, M+K$. 
Accordingly, we parametrize the potential $A$, as follows
\be
A_0  = \sum_{i=1}^{M} \( q_i P_i + r_i {P}_{-i} \) +
\sum_{a=M+1}^{M+K} U_a \,\, \a_a \cdot H^{(0)}
\lab{a20}
\ee
where $q_i$, $r_i$ and $U_a$ are fields of the model.

\subsubsection{The case $K=0$}

In this case, we have $\cgh = {\widehat {sl}} (M+1)$ and 
$\qs \equiv d$. The latter defines the homogeneous gradation. 
This example was discussed in detail in ref. \ct{AGZ95}.
The semisimple grade-one element $E$ is here given by
\be
E = \mu_M \cdot H^{(1)}
\ee
The kernel of ${\rm ad}\, E$ is :
\be
\cK= \kere = \{ {\widehat {sl}} (M) \oplus {\hat U}(1) \}
\ee
with ${\hat U}(1)$ being generated by $\mu_M \cdot H^{(k)}$, $k\in \IZ$
and defining the center of $\kere$ :
\be
\cC (\cK)= {\rm center}\, \kere = \{ \mu_M \cdot H^{(k)}\,\, , \,\,   k \in \IZ \}
\ee
Therefore, the dressing formalism associates the isospectral flow for 
each element :
\be
b_{k} \equiv \mu_M \cdot H^{(k)} \, , \, \qquad \mbox{\rm $k$ being a
positive integer}
\ee

The potential $A$ :
\be
A  = \sum_{i=1}^{M} \( q_i E_{(\a_{i} + \a_{i+1} + \ldots +\a_{M})}^{(0)} 
+ r_i E_{- (\a_{i} + \a_{i+1} + \ldots +\a_{M})}^{(0)}\)
\lab{a-homo}
\ee
lies in the complement $\cM$ of $\cK$ within $\cgh_0$.
Note, that $\cgh/\cK$ is now a symmetric space.
\lskip
{\bf Acknowledgements}
H.A. is partially supported by NSF (PHY-9820663),
J.F.G. and A.Z. are partially supported by CNPq and Fapesp (Brazil) and
E.N. and S.P. are partially supported by Bulgarian
NSF grant {\sl F-904/99}.
Also, H.A., E.N. and S.P. gratefully acknowledge support from NSF
grant {\sl INT-9724747}.

\end{document}